\def\e{{\rm e}}
\def\del{\partial}
\def\half{{1\over2}}

\def\abs#1{{\left|{#1}\right|}}
\def\vev#1{\langle #1 \rangle}
\def\del{\partial}
\def\dslash{\del\kern-0.55em\raise 0.14ex\hbox{/}}

\def\mbar{\bar m}
\def\vbar{\bar v}
\def\Vb{\bar V}
\def\btheta{\bar\theta}
\newcommand{\PRD}[3]{{\it Phys. Rev.} {\bf D{#1}} (19{#3}) {#2}}
\newcommand{\PRDM}[3]{{\it Phys. Rev.} {\bf D{#1}} (20{#3}) {#2}}

\newcommand{\NPB}[3]{{\it Nucl. Phys.} {\bf B{#1}} {#2} (19{#3})}
\newcommand{\NPBM}[3]{{\it Nucl. Phys.} {\bf B{#1}} (20{#2}) {#3}}
\newcommand{\PLB}[3]{{\it Phys. Lett.} {\bf {#1}B} (19{#3}) {#2}}
\newcommand{\PLBM}[3]{{\it Phys. Lett.} {\bf {#1}B} (20{#3}) {#2}}
\newcommand{\PLA}[3]{{\it Phys. Lett.} {\bf {#1}A} (19{#3}) {#2}}
\newcommand{\PTP}[3]{{\it Prog. Theor. Phys.} {\bf {#1}} (19{#3}) {#2}}
\newcommand{\PTPM}[3]{{\it Prog. Theor. Phys.} {\bf {#1}} (20{#3}) {#2}}
\newcommand{\ANN}[3]{{\it Ann. Phys. (N.Y.)} {\bf {#1}}, {#2} (19{#3})}

\newcommand{\MPL}[3]{{\it Mod. Phys. Lett.} {\bf A{#1}} (19{#3}) {#2}}
\newcommand{\MPLM}[3]{{\it Mod. Phys. Lett.} {\bf A{#1}} (20{#3}) {#2}}
\newcommand{\NCA}[3]{{\it Nuovo. Cim.} {\bf #1} (19{#3}) {#2}}
\newcommand{\hepth}[1]{{\tt [hep-th/#1]}}
\newcommand{\hepph}[1]{{\tt [hep-ph/#1]}}
\newcommand{\hmu}{\hat\mu}
\newcommand{\hnu}{\hat\nu}

\documentclass[12pt]{article}
\usepackage{graphicx}
\begin{document}
\baselineskip=18pt
%%%%%%%%%%%%%%%%%%%%%%%%%%%%
\begin{titlepage}
%%%%% PREPRINT NUMBERS %%%%%%
\begin{flushright}
%DIAS-STP-03-XX \\
KOBE-TH 03-03  \\
TIT/HEP-496    \\
OU-HET-444/2003
%hep-th/0305xxx
\end{flushright}
%%%%%%%%%%%%%%%%%%% TITLE %%%%%%%%%%%%%%%%%%
\begin{center}{\Large Phase Structures of
$SU(N)$ Gauge-Higgs Models\\ on \\ Multiply Connected Spaces}
\end{center}
%%%%%%%%%%%%%%%% AUTHORS %%%%%%%%%%%%%%%%%%%%%%%
\vspace{1cm}
\begin{center}
Hisaki {Hatanaka}$^{(a),}$
\footnote{E-mail: hatanaka@th.phys.titech.ac.jp}
Katsuhiko {Ohnishi}$^{(b),}$
\footnote{E-mail: ohnishi@phys.sci.kobe-u.ac.jp},
Makoto {Sakamoto}$^{(c),}$
\footnote{E-mail: sakamoto@phys.sci.kobe-u.ac.jp} and \\
Kazunori {Takenaga}$^{(d),}$
\footnote{E-mail: takenaga@het.phys.sci.osaka-u.ac.jp}
\end{center}
%%%%%%%%%%%%%%%%%%%%%%% AFFILIATION %%%%%%%%%%%%
\vspace{0.2cm}
\begin{center}
%\small
${}^{(a)}$ {\it Department of Physics,
Tokyo Institute of Technology, Tokyo 152-8551, Japan}
\\[0.2cm]
${}^{(b)}${\it Graduate School of Science and Technology, Kobe University,
\\
Rokkodai, Nada, Kobe 657-8501, Japan}
\\[0.2cm]
${}^{(c)}$ {\it Department of Physics, Kobe University,
Rokkodai, Nada, Kobe 657-8501, Japan}
\\[0.2cm]
%%%%
${}^{(d)}$ {\it Department of Physics, Osaka University, 
Toyonaka, Osaka 560-0043, Japan}
%%%%%
%${}^{(d)}$ {\it School of Theoretical Physics,
%Dublin Institute for Advanced Studies, \\
%10 Burlington Road, Dublin 4, Ireland}
%%%%%%%
\end{center}
%%%%%%%%%%%%%%%%%% ABSTRACT %%%%%%%%%%%%%%%%%%%%%%%
\vspace{1cm}
\begin{abstract}
We study an $SU(N)$ gauge-Higgs model
with $N_F$ massless fundamental fermions
on $M^3\otimes S^1$.
The model has two kinds of order parameters
for gauge 
symmetry breaking: the component gauge field for the $S^1$ direction
(Hosotani mechanism) and the Higgs field (Higgs mechanism).
We find that the model possesses three phases called Hosotani, Higgs and
coexisting phases for $N=$ odd, while for $N=$ even, the model has only
two phases, the Hosotani and coexisting phases.
The phase structure depends on
a parameter of the model 
and the size of the extra dimension.
The critical radius and the order of the phase transition are determined.
We also consider the case that the representation of
matter fields
under the gauge group is changed.
We find some models, in which there is only
%one phase, independent of
one phase independent of
parameters of the models as well as the size of the extra dimension.
\end{abstract}
\end{titlepage}
%%%%%%%%%%
%\tableofcontents
%%%%%%%%%%%%
\newpage
%%%%%%%%%%%%%%%% INTRODUCTION %%%%%%%%%%%%%%%
\section{Introduction}
%%%%%%%%%%%%%%%%%%%%%%%%%%%%%%%%%%%%%%%%%%%%%
%%%%%%%%%%%%%%%%%%%%%%%%%%%%%%%%%%%%%%%%%%%%%

Recently, physics with extra dimensions has been studied
extensively in connection with the long standing
problems, namely, new mechanism and/or the
origin of (gauge, super) symmetry breaking in elementary particle physics.
It can provide us new insight and understanding for low-energy physics.
In fact, it has been pointed out\cite{SUSY}
that new mechanism of spontaneous supersymmetry breaking
is possible in a certain class of models as a consequence
of the breakdown of the translational invariance
for the extra dimension $S^1$\cite{translation,O(N) model}.
Furthermore, one of our authors (M.S) and his
collaborators have shown\cite{monopole} that the rotational invariance of
$S^2$ is
spontaneously broken in a monopole background above some critical
radius due to the appearance of vortex configuration as vacuum
configuration. 
\par
%%%%%%%%
When one considers gauge-Higgs models on space-time with some of the
space directions being compactified on
a multiply connected space,
one should take account of gauge symmetry breaking
through the Hosotani mechanism\cite{Hosotani}.
The mechanism essentially occurs due to
quantum corrections in the extra dimension, reflecting the topology of
the compactified space. It is possible for the component gauge field for
the compactified direction to acquire nonvanishing vacuum expectation
values (VEV). The Hosotani mechanism has been studied extensively in
(supersymmetric) gauge
models\cite{HosotaniS2,HosotaniT2,HP,Hoso-ft,Pattern-Matter,Pattern-BC,HIL}
and, in particular, paid much attention in the context of orbifold
compactifications\cite{orbifold,Hoso-orbi}. On the other hand, the Higgs
mechanism also breaks gauge symmetry by the nonvanishing VEV for the
Higgs field even at the tree level. This suggests that if we consider
the gauge-Higgs model on such the space-time, the gauge symmetry can be
broken by both or either of the two mechanisms due to the existence of
the two kinds of 
the order parameters
for gauge symmetry breaking.  \par
%%%%%%%%
In a previous paper\cite{previous} we showed that phase
structures of gauge-Higgs models on $M^3\otimes S^1$ are
nontrivial, where $M^3 (S^1)$ is three-dimensional Minkowski
space-time (a circle). In the paper we studied the phase structure of the
simplest $SU(2)$ gauge-Higgs model and found three
different phases called Hosotani, Higgs and coexisting phases.
In each phase the VEVs for the two order parameters
take the different forms and values.
The structure depends on
a parameter 
of the model and the size of $S^1$.
The critical radius and the order of the phase transition
were determined explicitly. We also pointed out that the phase structure
could provide 
a new approach
to the gauge hierarchy problem in grand unified theory (GUT).
\par
%%%%%%%%
This paper is a generalization of the previous work. In particular, we
shall investigate the phase structure of
an $SU(N)$ gauge-Higgs model with
$N_F$ massless fermions
on $M^3\otimes S^1$. In the next section we analyze the phase structure
of the model, in which both the fermion and Higgs fields belong
to the fundamental representation under $SU(N)$.
We will find that the phase structure of the model is very
different, depending on
whether $N$ is even or odd. For the case $N=$ odd, there are the three
phases, and the structure is
similar to the one obtained
in the previous paper.
Only two phases, the Hosotani and the coexisting phases, appear
for $N=$ even,
and the Higgs phase does not exist for
finite sizes
of the extra dimension. We also determine the critical radius and
the order of the phase transition. In the models, the Hosotani mechanism
works as the restoration of the gauge symmetry.
In Sec. $3$, we consider the case that the representation of
matter fields
under the gauge group is changed. We find some models whose
phase structures do not depend on
the parameters of the models and also
the size of the extra dimension.
The final section is devoted to conclusions and discussions. 
Details of calculations will be given in Appendix.
%%%%%%%%%%%%%%%% Section 2 %%%%%%%%%%%%%%%%%%
%%%%%%%%%%%%%%%%%%%%%%%%%%%%%%%%%%%%%%%%%%%%%
\section{$SU(N)$ Gauge-Higgs Model}
%%%%%%%%%%%%%%%%%%%%%%%%%%%%%%%%%%%%%%%%%%%%%
%%%%%%%%%%%%%%%%%%%%%%%%%%%%%%%%%%%%%%%%%%%%%
We study the vacuum structure of
an $SU(N)$ gauge-Higgs model
with 
$N_F$ massless fundamental fermions.
The Higgs field also belongs to the fundamental
representation under $SU(N)$. We take our space-time to
be $M^3\otimes S^1$ in order to perform
analytic calculations, where $M^3$ and $S^1$ stand for
three-dimensional Minkowski space-time and a circle with
radius $R$, respectively. Our action is
%%%%%%
\begin{equation}
S=\int d^3x \int_0^Ldy\left(
-\half{\rm tr}F_{\hmu\hnu}F^{\hmu\hnu}+
\sum_{I=1}^{N_F}{\bar\psi}_I i\Gamma^{\hmu}D_{\hmu}\psi_I
+(D^{\hmu}\Phi)^{\dagger}D_{\hmu}\Phi - V(\Phi^{\dagger}, \Phi)\right),
\end{equation}
where the Higgs potential is given by
%%%%%%
\begin{equation}
V(\Phi^{\dagger}, \Phi)=-m^2\Phi^{\dagger}\Phi
+{\lambda\over 2}(\Phi^{\dagger}\Phi)^2.
\label{potential}
\end{equation} 
%%%%%
We have used a notation such as $x^{\hmu}\equiv (x^{\mu}, y)$ and
the length of the circumference of $S^1$ 
by $L=2\pi R$.
\par
%%%%%%%%%%%%%
As stated in the introduction, there are two kinds of the
order parameters
for gauge symmetry breaking.
One is the component gauge field $A_y$ for
the $S^1$ direction, which is related with the
Hosotani mechanism, and the mechanism is essentially caused by
quantum corrections
in the extra dimension. The other one
is the Higgs field, and the Higgs mechanism works
even at the tree-level.
Taking the order parameters into account,
we study the effective potential for $\vev{A_y}$ and $\vev{\Phi}$
parametrized by
%%%%%%%%
\begin{equation}
gL \vev{A_y}={\rm diag}(\theta_1, \theta_2, \cdots, \theta_N),~~
\vev{\Phi}={1\over\sqrt{2}}(v, 0, \cdots, 0)^T,
\label{vev}
\end{equation}
where $\sum_{i=1}^{N}\theta_i=0$ and $v$ is a real constant.
Here, we have arranged $\theta_i$ in such a way that
$\abs{\hat{\theta}_1}\leq \abs{\hat{\theta}_2}\leq
\cdots\leq\abs{\hat{\theta}_{N}}$,
where $\hat{\theta}_i = \theta_i$ mod $2\pi$ with
$\hat{\theta}_i \le \pi$.
This can be done without loss of generality.
We showed in the appendix that the parametrization
for the vacuum expectation value (VEV) of the Higgs field
given in Eq.(\ref{vev}) is enough to study the vacuum structure of the
model. 
We assume $N_F$ is so large
that the leading order correction comes from the fermion one-loop
correction to $\vev{A_y}$ alone. Then, the effective potential is given by
%%%%%%%%%%%%
\begin{eqnarray}
V &= &-\half m^2 v^2+
{\lambda\over 8}v^4+
{{\hat{\theta}_1^2v^2}\over {2L^2}}
+{A\over {\pi^2 L^4}}\sum_{n=1}^{\infty}\sum_{i=1}^{N}{1\over n^4}
\cos(n\theta_i)\label{effpotu}\\
&=&{1\over L^4}
\left(-\half \mbar^2 \vbar^2+
{\lambda\over 8}\vbar^4+
\half\hat{\theta}_1^2 \vbar^2
+{A\over {\pi^2}}\sum_{n=1}^{\infty}\sum_{i=1}^{N}{1\over n^4}
\cos(n\theta_i)\right)\equiv \Vb L^{-4},
\label{effpotd}
\end{eqnarray}
where $A\equiv 2^2 N_{F}$ and the number $2^2$ counts the physical
degrees of freedom of a Dirac fermion. Here, we have introduced
the dimensionless quantities, $\mbar\equiv mL, \vbar\equiv vL$.
The first two terms in Eq. (\ref{effpotu}) are nothing
but the classical Higgs potential, and the third term
comes from the interaction between the gauge and Higgs fields in
$D_y\Phi$, which, as we 
will see later, plays an important role to determine the phase structure
of the model. The fourth term stands for the one-loop correction
from the fermions.
We have neglected other
one-loop corrections arising from the gauge and Higgs fields
under the assumption that the number of
the fermions $N_F$ is sufficiently large
and also 
that the couplings $g$ and $\lambda$ are sufficiently small.
We make use of the assumption throughout the paper.
\par
%%%%%%%%%%%
If we look at the dependence of the effective potential on
the scale $L$ in Eq.(\ref{effpotu}), it suggests that
the vacuum structure changes according to
the size of the extra dimension. When $L$ is large enough, the quantum
correction in the extra dimension is suppressed and the leading order
contribution is given by the classical Higgs potential, so that
the Higgs field acquires the nonvanishing VEV.
The next leading order contribution, the third term
in Eq.(\ref{effpotu}),
yields vanishing $\hat{\theta}_1$ in order to minimize
the potential in the large $L$ limit.
On the other hand, if $L$ is small enough, the
quantum correction in the extra dimension dominates the effective
potential, and we would obtain
nonzero values
of $\theta_i$. 
Then, the next leading order term,
the third term, 
would enforce to result in $v=0$.
This simplified discussion implies that the vacuum structure depends on the
size of the extra dimension. One, of course, needs to study the
effective potential carefully in order to determine the vacuum structure
of the model.
\par
%%%%%%%%%%  
Let us now study the vacuum structure of the model.
We follow the standard procedure to
find the vacuum configuration. We first solve equations of
the first derivative of the effective potential (\ref{effpotd})
with respect to the order parameters,
%%%%%%%%%%%%%
\begin{eqnarray}
{{\del \Vb}\over{\del\vbar}}&=&\vbar \left(-\mbar^2+{\lambda\over 2}\vbar^2
+\hat{\theta}_1^2\right)=0,
\label{deru}\\
%%%%%%%
{{\del \Vb}\over{\del\theta_1}}&=&
\hat{\theta}_1 \vbar^2
+{A\over \pi^2}\sum_{n=1}^{\infty}
{{-1}\over n^3}\left(\sin(n\theta_1)+\sin(n\sum_{i=1}^{N-1}\theta_i)
\right)=0,\label{derd}\\
%%%%%%%%%%%
{{\del \Vb}\over{\del\theta_k}}&=&
{A\over \pi^2}\sum_{n=1}^{\infty}{{-1}\over n^3}
\left(\sin(n\theta_k)+\sin(n\sum_{i=1}^{N-1}\theta_i)\right)=0,
\quad k=2, \cdots, N-1,
\label{dert}
\end{eqnarray}
%%%%%%%%%%
where we have used $\theta_N = - \sum_{i=1}^{N-1}\theta_i$.
Solutions to the equations are
candidates 
of the vacuum configuration.
Then, we analyze the stability of
the solutions
against small fluctuations, and this constrains
allowed regions of the solutions
as the local minimum of the effective potential. Among various, if any,
candidates of those configurations, the global minimum of the effective
potential is given by the configuration which gives the lowest energy.
Following these steps, one can obtain the vacuum structure of the model.
\par
%%%%%%%%%
The equation (\ref{deru}) leads to
%%%%%%%
\begin{eqnarray}
&&\vbar =0,\label{hvevu}\\
&{\rm or}&\nonumber\\
&&-\mbar^2+{\lambda\over 2}\vbar^2+\hat{\theta}_1^2=0.
\label{hvevd}
\end{eqnarray}
%%%%%%
For the first case (\ref{hvevu}), the equations (\ref{derd}) and
(\ref{dert}) are unified into an equation,
%%%%%%%%%%%
\begin{equation}
{A\over \pi^2}\sum_{n=1}^{\infty}{{-1}\over n^3}
\left(\sin(n\theta_k)+\sin(n\sum_{i=1}^{N-1}\theta_i)\right)=0,
\quad k=1, \cdots, N-1.
\label{eqhosotani}
\end{equation}
We call the solution to this equation type I, and the solution
describes the Hosotani phase. On the other hand, for the
second case (\ref{hvevd}) we solve the coupled equations,
%%%%%%%%%
\begin{eqnarray}
{2\over\lambda}(\mbar^2 -\hat{\theta}_1^2)\hat{\theta}_1
+{A\over \pi^2}\sum_{n=1}^{\infty}
{{-1}\over n^3}\left(\sin(n\theta_1)+\sin(n\sum_{i=1}^{N-1}\theta_i)
\right)&=&0,\label{eqcoexistu}\\
%%%%%%%%%%%
{A\over \pi^2}\sum_{n=1}^{\infty}{{-1}\over n^3}
\left(\sin(n\theta_k)+\sin(n\sum_{i=1}^{N-1}\theta_i)\right)&=&0,
\quad k=2, \cdots, N-1.
\label{eqcoexistd}
\end{eqnarray}
A solution to the equation is called type II or type III, depending on
whether the solution has the scale dependence on the extra dimension or not.
The type II (III) corresponds to the Higgs (coexisting) phase,
whose vacuum expectation values are independent of (dependent on)
the scale of the extra dimension.
In order to avoid unnecessary complexity,
details of
calculations to solve these equations will be given in the appendix.
We find that it is convenient to discuss the vacuum
structure separately, depending on
whether $N$ is odd or even.
Let us first study the case $N=$ odd.
%%%%%%%%%%%%%%% Section 2.1 %%%%%%%%%%%%%%%%%
%%%%%%%%%%%%%%%%%%%%%%%%%%%%%%%%%%%%%%%%%%%%%
\subsection{$N=$ odd}
%%%%%%%%%%%%%%%%%%%%%%%%%%%%%%%%%%%%%%%%%%%%%
%%%%%%%%%%%%%%%%%%%%%%%%%%%%%%%%%%%%%%%%%%%%%
There are three types of possible vacuum configurations, as shown
in the appendix,
%%%%%%%%%%%
\begin{eqnarray}
{\rm type~~I}&\cdots&\left\{\begin{array}{l}
gL\vev{A_y}={\rm diag}\left({{N-1}\over N}\pi,
\cdots,{{N-1}\over N}\pi, -{{(N-1)^2}\over N}\pi\right),\\[0.3cm]
\vev{\Phi}={1\over\sqrt{2}}(0,\cdots, 0)^T,
\end{array}\right. \label{hosotani}\\[0.3cm]
%%%%%%%%%%
{\rm type~~II}&\cdots&\left\{\begin{array}{l}
gL\vev{A_y}={\rm diag}\left(0, \pi, \pi, \cdots, -(N-2)\pi\right),\\[0.3cm]
\vev{\Phi}={1\over\sqrt{2}}(\sqrt{2\over\lambda}m,0,\cdots, 0)^T,
\end{array}\right.\label{higgs}\\[0.3cm]
%%%%%%%%%%%
{\rm type~~III}&\cdots&\left\{\begin{array}{l}
gL\vev{A_y}={\rm diag}\left(\theta_1^-, \pi-{\theta_1^-\over{N-1}},\cdots,
\pi-{\theta_1^-\over{N-1}},
-((N-2)\pi+{\theta_1^-\over{N-1}})\right),\\[0.3cm]
\vev{\Phi}={1\over\sqrt{2}}(v,\cdots, 0)^T.
\end{array}\right.\label{coexist}
\end{eqnarray}
The type III solution depends on the
scale $\mbar$, 
and accordingly, 
$\vbar=\sqrt{{2\over\lambda}(\mbar^2-(\theta_1^{-})^2)}$ does as well.
The explicit form of $\theta_1^-(\mbar)$ is given by Eq.(\ref{cosol}) in
the appendix. As we stated before, we call the
vacuum configuration corresponding to
the type I, II and III solutions
the Hosotani, Higgs and coexisting phases, respectively.
\par
%%%%%%%%%%
Given the vacuum configuration, the
gauge symmetry in each phase is generated by the
generators $T^a$ of $SU(N)$ which commute with the Wilson line,
%%%%%%%%%%
\begin{equation}
W\equiv {\cal P}{\rm exp}\left(ig\oint_{S^1}dy \vev{A_y}\right)
={\rm diag}\left(\e^{i\theta_1}, \e^{i\theta_2}, \cdots, \e^{i\theta_N}
\right)\quad
{\rm and} \quad T^a\vev{\Phi}=0.
\end{equation}
Let us note that the phase $\theta_i$ is
defined modulo $2\pi$.
It is easy to observe that the $SU(N)$ gauge
symmetry is not broken in the Hosotani phase because the Wilson
line for the configuration (\ref{hosotani}) is proportional to the identity
matrix, $W={\rm exp}(i({{N-1}\over N}\pi)){\bf 1}_{N\times N}$.
In the Higgs phase, the $SU(N)$ gauge symmetry
is broken to $SU(N-1)\times U(1)$ by the Wilson line and the $U(1)$
symmetry is broken by the Higgs VEV, so that the
residual gauge symmetry is $SU(N-1)$.
Likewise, in the coexisting phase, the residual gauge symmetry is $SU(N-1)$.
\par
%%%%%%%%%%%%
It is important to note that each type of the vacuum configuration
has the restricted region determined by the
scale $\mbar$, in which the configuration is stable
against small fluctuations.
The region also depends on the parameter $t\equiv \lambda A=4\lambda N_F$.
Let us quote relevant results from the appendix that are
necessary to determine the phase structure of the model.
The Hosotani phase (type I) is stable for
%%%%%%%
\begin{equation} 
0 < \mbar < {{N-1}\over N}\pi\equiv \mbar_2,
\label{regionu}
\end{equation}
%%%%%%%%%
and the Higgs phase (type II) is stable when $\mbar$ satisfies
%%%%%%%%
\begin{equation} 
\mbar > \left({{2N-3}\over{N-1}}{t\over{24}}\right)^{\half}\equiv
\mbar_3.
\label{regiond}
\end{equation}
%%%%%%%%%%%%%
In the coexisting phase (type III), $\theta_1^-(\mbar)$ must satisfy
the reality condition $(\theta_1^-({\mbar}))^*=\theta_1^-(\mbar) $ and
%%%%%%%%%
\begin{equation} 
0 \leq \theta_1^-({\mbar})\leq {{N-1}\over N}\pi .
\label{restrictodd}
\end{equation} 
These requirements on $\theta_1^-(\mbar)$
restrict the allowed region of the coexisting phase
in the parameter space of $(\bar{m}, t)$.
The analysis in the appendix shows that the coexisting phase
lies in the region
%%%%%%%%%
\begin{eqnarray} 
\mbar_2 \leq \mbar \le \mbar_3 && \quad{\rm for}\quad
t\ge 48\pi^2\frac{(N-1)^3}{N(N^2-3)} ,
\label{regionofcoexist1}\\
\mbar_1 \leq \mbar \le \mbar_3 && \quad{\rm for}\quad
t < 48\pi^2\frac{(N-1)^3}{N(N^2-3)} ,
\label{regionofcoexist2}
\end{eqnarray} 
where $\bar{m}_1$ is the critical scale above which the realty
condition is furnished and is given by Eq.(\ref{m_1}) in the
appendix.
\par
%%%%%%%%%
Now, we are ready to determine the phase structure of the model.
As shown in Fig.1, the lines $\bar{m}_i (i=1,2,3)$ divide the
$\bar{m}$-$t$ plane into the several regions. 
%%
%pieces.
%%%
%Some of them allow only one phase, which must be the vacuum
%configuration.
%%%
Some region allows only one phase, which is nothing but the
vacuum configuration. There are, however, overlapping regions
in which two of the three phases remain as candidates of the 
vacuum configuration. In this case, one has to determine which 
phase gives the lowest energy among them. Fig.1 will help us 
understand the phase structure of the model. 
\par
%%%%%%%%%
Since $\bar{m}_1=\bar{m}_2$ at $t=t_1\equiv 
48\pi^2\frac{(N-1)^3}{N(N^2-3)}$
and $\bar{m}_2=\bar{m}_3$ at $t=t_2\equiv 24\pi^2\frac{(N-1)^3}{N^2(2N-3)}$
(no other intersections of the curves $\bar{m}_i$ for $t>0$),
it is convenient to consider separately the three parameter
regions of $t$:
%%%%%%%%%
\begin{eqnarray}
48\pi^2{{(N-1)^3}\over{N(N^2-3)}} < & t, &
\hspace{3.9cm}(\mbar_2 < \mbar_3) ,\label{cou}\\
24\pi^2{{(N-1)^3}\over {N^2(2N-3)}}
< &t & \leq  48\pi^2{{(N-1)^3}\over{N(N^2-3)}},~~~
(\mbar_1 \leq \mbar_2 < \mbar_3),\label{cod}\\
& t & \leq 24\pi^2{{(N-1)^3}\over {N^2(2N-3)}},~~
(\mbar_1 < \mbar_3 \leq \mbar_2).\label{cot}
\end{eqnarray}
Here, the relative magnitude of $\bar{m}_i$ for each parameter
region of $t$ is shown in the parenthesis.\footnote{
We do not need to take $\bar{m}_1$ into account for the first
case (\ref{cou}) in our analysis.
}
\par
%%%%%%%%
\begin{flushleft}
(i)~~$t > 48\pi^2{{(N-1)^3}\over{N(N^2-3)}}$
\end{flushleft}
We immediately observe that the
scale $\mbar_2$ $(\mbar_3)$ is the phase
boundary between the Hosotani phase and
the coexisting one (the coexisting phase and the Higgs one).
There is no overlapping region of the phases
for this parameter region of $t$. Thus, the
vacuum configuration is given by
%%%%%%%
\begin{equation}
\mbox {vacuum configuration}=\left\{\begin{array}{lll}
\mbox{Hosotani phase} &\mbox{for}& \mbar < \mbar_2,\\
\mbox{coexisting phase} &\mbox{for}& \mbar_2 < \mbar < \mbar_3,\\
\mbox{Higgs phase} & \mbox{for}& \mbar_3 < \mbar.
\end{array}\right.
\label{solu}
\end{equation}
%%%%%
The order parameters in the Hosotani and coexisting
phases 
(the coexisting and Higgs phases)
are connected continuously
at the phase boundary $\mbar_2$ $(\mbar_3)$, so that the phase
transition is the second order.
%%%%%%%%%%%
\begin{flushleft}
(ii)~~$24\pi^2{{(N-1)^3}\over {N^2(2N-3)}}
< t \leq  48\pi^2{{(N-1)^3}\over{N(N^2-3)}}$
\end{flushleft}
For this parameter region of $t$, the Hosotani and coexisting phases
overlap
between $\mbar_1$ and $\mbar_2$. Let us consider
the quantity $\Delta\Vb\equiv\Vb_{Hosotani}-\Vb_{coexisting}$,
which is a monotonically increasing function with respect to $\bar{m}$,
%%%%%%%%
\begin{equation}
{{\del\Delta\Vb}\over{\del\mbar^2}}=\half \vbar^2(\mbar)\geq 0,
\label{monou}
\end{equation}
as shown in the appendix. By denoting the scale
giving $\Vb_{coexisting}=\Vb_{Hosotani}$ by $\mbar_4$,
we can conclude that for $\mbar \leq \mbar_4$
$(\mbar_4 < \mbar\leq \mbar_3)$, the Hosotani (coexisting) phase
is realized as the vacuum configuration. The Higgs phase can exist
for $ \mbar_3 < \mbar$. Thus, we obtain that
%%%%%%%%%%%
\begin{equation}
\mbox {vacuum configuration}=\left\{\begin{array}{lll}
\mbox{Hosotani phase} &\mbox{for}& \mbar < \mbar_4,\\
\mbox{coexisting phase} &\mbox{for}& \mbar_4 < \mbar < \mbar_3,\\
\mbox{Higgs phase} & \mbox{for}& \mbar_3 < \mbar.
\end{array}\right.
\label{sold}
\end{equation}
The explicit form of the critical scale $\bar{m}_4$ is given by
%%%%%%%%%%%
\begin{eqnarray}
\bar{m}_4 = 
2\pi\sqrt{\frac{1}{2a}\left( b + \sqrt{b^2 - 4ac}\right)} ,
\label{m_4}        
\end{eqnarray}
where
%%%%%%%%%%%
\begin{eqnarray}
a &=& 48 N^4 (N^2 -3N+3)^2 \biggl( 3(N-1)^3 +
      2N(N^2-3N+3)\frac{t}{16\pi^2} \biggr) ,\label{m_4a}\\
b &=& 24 N^2 (N-1)^2 \biggl( 3(N-1)^3 (N^2-N-1)(3N^2-5N+1)\nonumber\\
   & &\ \    + 2N^3 (2N-3)^2 (N^2-3N+3) \frac{t}{16\pi^2}\biggr) ,
      \label{m_4b}\\
c &=& (N-1)^3 \biggl( -9(N-1)^4 (N^2-N-1)^2
      - 6N(N-1)(N^2-N-1)\nonumber\\
   & &\ \   \times(11N^4-36N^3+33N^2-9)\frac{t}{16\pi^2}
      + 4N^2(N^2-3)^3 \biggl( \frac{t}{16\pi^2}\biggr)^2
      \biggr) . \label{m_4c}
\end{eqnarray}
The phase transition at $\mbar=\mbar_3$
is the second order, while that at $\mbar=\mbar_4$ is the
first order because the order parameters are
not connected continuously.
%%%%%%%%%%%%
\begin{flushleft}
(iii)~~$ t \leq 24\pi^2{{(N-1)^3}\over {N^2(2N-3)}}$
\end{flushleft}
Let us first compare the potential energy of the Hosotani phase with that
of the Higgs phase. The scale $\mbar_5$ is the critical scale, at which
$\Vb_{Hosotani}=\Vb_{Higgs}$ holds. Then, as shown in the appendix, we
obtain that 
%%%%%%%%
\begin{eqnarray}
\Vb_{Hosotani}< \Vb_{Higgs}\qquad {\rm for}
\qquad \mbar < \mbar_5,\label{hhu}\\
\Vb_{Hosotani}> \Vb_{Higgs}
\qquad {\rm for}\qquad \mbar > \mbar_5,
\label{hhd}
\end{eqnarray} 
where
%%%%%%%%
\begin{equation}
\mbar_5\equiv \left({{(N-1)(N^2-N-1)}\over N^3}
{\pi^2\over{24}}t\right)^{1\over 4}.
\label{regiont}
\end{equation}
The parameter region of $t$ is further classified into two
cases,
depending on the relative magnitude between $\mbar_5$ and $\mbar_3$:
%%%%%%%%
\begin{flushleft}
(iii-a)~~$24\pi^2 {{(N-1)^3(N^2-N-1)}\over {N^3 (2N-3)^2}}
< t \leq 24\pi^2{{(N-1)^3}\over {N^2(2N-3)}}$
\end{flushleft}
In this case, the relative magnitude of $\bar{m}_i\ (i=1,\cdots,5)$
is given by $\bar{m}_1<\bar{m}_4\le\bar{m}_5\le\bar{m}_3<\bar{m}_2$
(see Fig.1).
It immediately follows that the vacuum configuration
is uniquely determined as
%%%%%%%%%%%
\begin{equation}
\mbox {vacuum configuration}=\left\{\begin{array}{lll}
\mbox{Hosotani phase} &\mbox{for}& \mbar < \mbar_4,\\
\mbox{coexisting phase} &\mbox{for}& \mbar_4 < \mbar < \mbar_3,\\
\mbox{Higgs phase} & \mbox{for}& \mbar_3 < \mbar.
\end{array}\right.
\label{solt}
\end{equation}
The vacuum structure is similar to the case (ii), and
the phase transition at $\bar{m}=\bar{m}_3$ ($\bar{m}=\bar{m}_4$)
is the second (first) order.
%%%%%%%%%%%%%%
\begin{flushleft}
(iii-b)~~$ t\leq 24\pi^2 {{(N-1)^3(N^2-N-1)}\over {N^3 (2N-3)^2}}$
\end{flushleft}
In this case, 
we have $\mbar_1 \le \mbar_3 \le \bar{m}_5 < \bar{m}_2$ (see Fig.1).
We observe that the Hosotani and coexisting phases
overlap
between $\mbar_1$ and $\mbar_3$. Let us recall that the difference
of the potential energy between the Hosotani phase and the coexisting
one, $\Delta\Vb$, is
the monotonically increasing function with respect to $\bar{m}$,
and we find that 
%%%%%%%%%%
\begin{equation}
\Delta \Vb(\mbar=\mbar_3)={t\over{2\lambda}}
\left({{2N-3}\over{24(N-1)}}\right)^2
\left(t- 24\pi^2 {{(N-1)^3(N^2-N-1)}\over {N^3 (2N-3)^2}}\right) \leq 0,
\label{valu}
\end{equation}
%%%%%%%%%%%
{\it i.e.}  $\Vb_{Hosotani} \leq \Vb_{coexisting}$ for the parameter
region of $t$ 
under consideration.
This implies that there is no coexisting phase for this parameter
region of $t$.
Thus, taking Eq.(\ref{hhd}) into account,
we obtain that
%%%%%%%%%%%
\begin{equation}
\mbox {vacuum configuration}=\left\{\begin{array}{lll}
\mbox{Hosotani phase} &\mbox{for}& \mbar < \mbar_5,\\
\mbox{Higgs phase} & \mbox{for}& \mbar > \mbar_5.
\end{array}\right.
\label{solq}
\end{equation}
The order parameters are not connected continuously at $\mbar_5$.
The phase transition between the two phases is the
first order
and there is no coexisting
phase for this parameter region of $t$.
\par
%%%%%%%%%%
Collecting all the results obtained above, we depict the phase
structure of the model in Fig.2. It should be noted that the Hosotani
mechanism, which is usually known to break down gauge symmetry, provides
a mechanism of the {\it restoration}
of the gauge symmetry in the model.
%\newpage
%%%%%%%%%%%%HERE COMES FIG 1 %%%%%%%%
%\begin{center}
%\fbox{Fig. $1$}
%\end{center} 
%%%%%%%%%%%%%%%%%%%%%%%%%%%%%%%%%%%%%%%%%%%%%
%%%%%%%%%%%%%%% N=EVEN PART %%%%%%%%%%%%%%%%%
%%%%%%%%%%%%%%%%%%%%%%%%%%%%%%%%%%%%%%%%%%%%%
\subsection{$N=$ even $\geq 4$}
%%%%%%%%%%%%%%%%%%%%%%%%%%%%%%%%%%%%%%%%%%%%%
%%%%%%%%%%%%%%%%%%%%%%%%%%%%%%%%%%%%%%%%%%%%%
Let us study the case $N=$ even $(\geq 4)$. The type I solution
corresponding to the Hosotani phase is given by solving
Eq.(\ref{eqhosotani}).
We obtain, as shown in the appendix, that
%%%%%%%%%%%
\begin{equation}
{\rm type~~I}~~\cdots~~\left\{\begin{array}{l}
gL\vev{A_y}={\rm diag}\left(\pi, \pi,\cdots,
\pi, -(N-1)\pi\right),\\[0.3cm]
\vev{\Phi}={1\over\sqrt{2}}(0,0,\cdots, 0)^T,
\end{array}\right.
\label{hosotanid}
\end{equation}
where the phase is stable for the region given by
%%%%%%
\begin{equation}
0<\mbar < \pi.
\label{regionq}
\end{equation}
The Wilson line for the configuration (\ref{hosotanid})
is $-{\bf 1}_{N\times N}$ and commutes with all the generators
of $SU(N)$, so that the $SU(N)$ symmetry is not broken in the phase.
\par
%%%%%%%%%%%%
In order to investigate other phases,
one has to 
solve the third order equation with respect to $\theta_1$, 
%%%%
%as shown in the appendix,
%%%%%%%%%
\begin{equation}
\left(1+{{t\alpha}\over{24\pi^2}}\right)\theta_1^3
-\left({{t\alpha}\over{8\pi}} +6\pi\right)\theta_1^2
+\left({{t\beta}\over{24}}-\mbar^2+12\pi^2\right)\theta_1
+{{t\gamma}\over{24}}\pi +2\pi\mbar^2 -8\pi^3=0,
\label{eqevendt}
\end{equation}
where $\alpha, \beta$ and $\gamma$ are constants
and any solution to Eq.(\ref{eqevendt}) has to lie in the range
%%%%%%%%%
\begin{eqnarray}
\pi \le \theta_1 \le 2\pi ,
\label{restricteven}
\end{eqnarray}
as shown in the appendix.
This equation has the very different structure
from that for $N=$ odd (See Eq.(\ref{cohiggsu}) in the appendix).
It does not have the solution
of $\theta_1=2\pi$ for any values of $t$ and $\mbar$.
This implies that the Higgs phase
%%%%%%%%%%
\begin{equation}
{\rm type~~II}~~\cdots~~\left\{\begin{array}{l}
gL\vev{A_y}={\rm diag}\left(0, {{N-2}\over{N-1}}\pi,\cdots,
{{N-2}\over{N-1}}\pi, -{{(N-2)^2}\over{N-1}}\pi\right),\\[0.3cm]
\vev{\Phi}={1\over\sqrt{2}}(\sqrt{2\over\lambda}m,0,\cdots, 0)^T,
\end{array}\right.
\label{higgsd}
\end{equation}
does not exist unlike the case $N=$ odd.
The Higgs phase can be realized only
in the limit 
$L \rightarrow \infty$ (or $\bar{m}\rightarrow \infty$),
as we will see later. The $SU(N)$ gauge symmetry is broken 
to $SU(N-1)\times U(1)$ by the Wilson line and the Higgs VEV
breaks the $U(1)$, so that the residual gauge symmetry 
is $SU(N-1)$ for the configuration (\ref{higgsd}). 
\par
%%%%%%%%%%
In order to see that the vacuum configuration is expected to
approach the Higgs phase in the limit $\bar{m}\rightarrow \infty$,
let us first note that the classical Higgs potential dominates
the effective potential in the limit.
Then, we obtain the nonvanishing Higgs VEV $v=\sqrt{2m/\lambda}$.
It follows that the equation (\ref{deru}) results in
$\hat{\theta}_1 =0$ or $\theta_1 =0$ mod $2\pi$.
%%%%%
%This result will be also derived from Eq.(\ref{eqevendt})
%by taking the limit $\bar{m}\rightarrow \infty$.
%%%%%
This result is also derived from Eq.(\ref{eqevendt})
by taking the limit $\bar{m}\rightarrow \infty$.
For these values of the order parameters, the equation we have
to solve becomes the same equation as the one that produces
the Hosotani phase for the case $N=$ odd, but in the present
case, $N$ is replaced by $N-1$ ($=$ odd)
with the nonvanishing $\bar{v}$.
Hence, we finally arrive at the solution (\ref{higgsd}).
\par
%%%%%%%%%%
Our task is now to solve the equation (\ref{eqevendt}) for finite
sizes of $S^1$ and to confirm the phase structure for the
case $N=$ even depicted in the Fig.3.
The coexisting phase is given by
%%%%%%%%%%
\begin{equation}
{\rm type~~III}~~\cdots~~\left\{\begin{array}{l}
gL\vev{A_y}={\rm diag}\left(\theta_c, {N\over{N-1}}\pi
-{\theta_c\over{N-1}},\cdots,{N\over{N-1}}\pi-{\theta_c\over{N-1}},
{{-\theta_c}\over{N-1}}-{{N(N-2)}\over{N-1}}\pi\right),\\[0.3cm]
\vev{\Phi}={1\over\sqrt{2}}(\sqrt{2\over\lambda}v,0,\cdots, 0)^T,
\end{array}\right.
\label{coexistd}
\end{equation}
where $\vbar =\sqrt{{2\over\lambda}(\mbar^2-(2\pi-\theta_c)^2)}$ and
$\theta_c$ is the solution for the coexisting phase (see below).
The $SU(N)$ gauge symmetry is broken to $SU(N-1)\times U(1)$ by the
Wilson line and the Higgs VEV breaks the $U(1)$, so that the
residual gauge symmetry is $SU(N-1)$ for the vacuum
configuration (\ref{coexistd}).
%%%%%%%%%%HERE COMES FIG.2 %%%%%%%%%
%\begin{center}
%\fbox{Fig.$2$}
%\end{center}
%%%%%%
\par
%%%%%%%%%%%%%%%
In order to confirm that the phase structure is actually given
by the Fig.3, it is convenient to consider intersections of
two functions defined by
%%%%%%%%%%
\begin{eqnarray}
F(\theta_1)&\equiv &2(\mbar^2-(2\pi - \theta_1)^2)(2\pi -\theta_1),\\
G(\theta_1)& \equiv &{{-t}\over{12\pi^2}}
\left(\alpha\theta_1^3 -3\pi \alpha\theta_1^2 +\beta \pi^2\theta_1
+ \gamma \pi^3\right).
\label{fgeq}
\end{eqnarray}
Let us note that $F(\theta_1)=G(\theta_1)$, in fact, reproduces
Eq.(\ref{eqevendt}) and $G(\theta_1)$ is dependent of $\bar{m}$.
Since the intersections of the functions $F(\theta_1)$ and
$G(\theta_1)$ have different behavior
for $t > 48\pi^2 (N-1)/N$ and $t < 48\pi^2 (N-1)/N$,
as discussed in the appendix, it is
convenient to investigate separately the phase structure for each region
of $t$ (see Figs.4 and 5).
%%%%%%%%%%
%\begin{center}
%\fbox{Figs. $3, 4$}
%\end{center}
%%%%%%%%%
\par
%%%%%%%%%%%%
\begin{flushleft}
(i)~~$t> 48\pi^2 {{N-1}\over N}$
\end{flushleft}
In this parameterization of $t$, there is one solution denoted by
$\theta_c$ for $\bar{m} > \pi$, as shown in Fig.4.
The solution satisfies the condition (\ref{restricteven}) and
is found to be stable, as discussed in the appendix.
Since the Hosotani phase is unstable for $\bar{m}>\pi$,
the coexisting phase must be the vacuum configuration for
$\bar{m}>\pi$.
\par 
%%%%%%%%%%%
When the scale $\bar{m}$ approaches $\pi$, $\theta_c$ becomes
closer to $\pi$ and is finally identical to $\pi$ at $\bar{m}=\pi$.
This implies that the type III solution (coexisting phase) becomes
identical to the type I solution (Hosotani phase).
As the scale becomes smaller than $\pi$, the solution is outside
of the required region (\ref{restricteven}).
Hence, there is no coexisting phase for $\bar{m}<\pi$,
so that the Hosotani phase must be the vacuum configuration
for $\bar{m}<\pi$.
Thus,
we obtain that  
%%%%%%%%%
\begin{equation}
\mbox {vacuum configuration}=\left\{\begin{array}{lll}
\mbox{Hosotani phase} &\mbox{for}& \mbar < \pi,\\
\mbox{coexisting phase} &\mbox{for}& \mbar > \pi.
\end{array}\right.
\label{vaceven1}
\end{equation}
Since the order parameters are connected continuously at the
phase boundary $\bar{m}=\pi$, the phase transition is
the second order.
%%%%%%%%%  
\begin{flushleft}
(ii)~~$t < 48\pi^2 {{N-1}\over N}$
\end{flushleft}
%%%%%%%%%%%%%%%%%
In this parameter region of $t$, we observe in Fig.5 that
there is one solution denoted by $\theta_c$ for $\bar{m}>\pi$.
The solution satisfies the condition (\ref{restricteven}) and is
stable, so that the coexisting phase is the vacuum configuration
for $\bar{m}>\pi$, as in the case (i).
\par
%%%%%%%%%%
Unlike the case (i), $\theta_c$ does not approach $\pi$ as
$\bar{m}\rightarrow \pi$.
When the scale $\bar{m}$ is equal to $\pi$, there appears a new solution
denoted by $\theta_{c}^{\prime} = \pi$, while the $\theta_c$
still lies between $\pi$ and $2\pi$, as shown in Fig.5.
If we go to smaller scales than $\pi$, there are two solutions
$\theta_{c}^{\prime}$ and $\theta_c$ with
$\pi<\theta_{c}^{\prime}\le \theta_c <2\pi$ for
$\bar{m}_1^{\prime} \le \bar{m} < \pi$, where the two solutions coincide at
$\bar{m}=\bar{m}_1^{\prime}$ (see Fig.5).
Since there are no solutions in the required region of $\theta_1$
below the scale $\bar{m}_1^{\prime}$, the coexisting phase disappears and
the Hosotani phase must be the vacuum configuration for
$\bar{m}<\bar{m}_1^{\prime}$.
\par
%%%%%%%%%%%%
One has to take care about what is happening  in the region
of $\mbar_1^{\prime} \leq \mbar \leq \pi$.
For this region, there are two solutions, $\theta_c$ and
$\theta_{c}^{\prime}$, to Eq.(\ref{eqevendt}) or the equation
$F(\theta_1)=G(\theta_1)$.
It turns out that the solution $\theta_c$ is stable but the other
one is unstable, as shown in the appendix.
Thus, there are two candidates for the vacuum configuration,
{\it i.e.} the Hosotani phase and the coexisting phase given by
$\theta_c$.
Since $\theta_{c}^{\prime}=\theta_c$ ($\theta_{c}^{\prime}=\pi$)
at $\bar{m}=\bar{m}_1^{\prime}$ ($\bar{m}=\pi$), we find that
the unstable solution $\theta_{c}^{\prime}$ becomes identical to
the coexisting phase given by $\theta_c$ (the Hosotani phase)
at $\bar{m}=\bar{m}_1^{\prime}$ ($\bar{m}=\pi)$.
This shows that the coexisting phase (the Hosotani phase) is not
the vacuum configuration at the boundary $\bar{m}=\bar{m}_1^{\prime}$
$(\bar{m}=\pi)$.
This observation implies that there exists a critical scale
$\bar{m}_4^{\prime}$ such that\footnote{
Although we can give an analytic expression for $\bar{m}_4^{\prime}$,
it will not be useful for practical purposes.
}
%%%%%%%%%%%%
\begin{equation}
\Vb(\theta_i, \vbar,\mbar_{4}^{\prime})\bigg|_{\rm type~I}
=\Vb(\theta_i, \vbar,\mbar_{4}^{\prime})\bigg|_{\rm type~III(\theta_c)},
\end{equation}
at which the first-order phase transition must occur.
Since the above equation is satisfied
only once for $\bar{m}_1^{\prime}\le\bar{m}\le\pi$,
we obtain the phase structure
for $t<48\pi^2\frac{N-1}{N}$ as
%%%%%%%%%
\begin{equation}
\mbox {vacuum configuration}=\left\{\begin{array}{lll}
\mbox{Hosotani phase} &\mbox{for}& \mbar < \bar{m}_{4}^{\prime},\\
\mbox{coexisting phase} &\mbox{for}& \mbar > \bar{m}_{4}^{\prime}.
\end{array}\right.
\label{vaceven2}
\end{equation}
\par
%%%%%%%%%
Collecting all the discussions we have made in this subsection,
we confirm the phase structure depicted in Fig.3.
We should emphasize again that the Hosotani mechanism works
as the restoration of the gauge symmetry, as in the case
$N=$ odd.
%%%%%%%%%%%%%%%%%%%%%%%%%%%%%%%%%%%%%%%%%%%%%%%
%%%%%%%%%%%%%%%%%%%%%%%%%%%%%%%%%%%%%%%%%%%%%%%
\section{Other Models}
%%%%%%%%%%%%%%%%%%%%%%%%%%%%%%%%%%%%%%%%%%%%%%%
%%%%%%%%%%%%%%%%%%%%%%%%%%%%%%%%%%%%%%%%%%%%%%%
In this section we study how phase structures change if we consider
different representations of matter fields
under the gauge group.
Let us first introduce $N_F$ fermions
in the adjoint representation
under $SU(N)$ instead of
those in the fundamental representation. Then, the last term
in Eq.(\ref{effpotu}), which
stands for the fermion 
one-loop correction, is replaced by
%%%%%%%
\begin{equation}
{A\over{\pi^2 L^4}}\sum_{n=1}^{\infty}\sum_{i,j=1}^{N}
{1\over n^4}\cos(n(\theta_i -\theta_j)).
\label{adjfermi}
\end{equation}
It has been known that
%this potential 
the function (\ref{adjfermi})
is minimized by the configuration
that breaks the $SU(N)$ gauge symmetry to $U(1)^{N-1}$
\cite{adjointmodel},
%%%%%%%%%
\begin{equation}
gL\vev{A_y}={\rm diag}\left({{N-1}\over N}\pi, {{N-3}\over N}\pi,
\cdots,
-{{N-3}\over N}\pi, -{{N-1}\over N}\pi \right).
\label{configadj}
\end{equation}
Note that a zero eigenvalue located at the
${{N+1}\over 2}$th component appears for the case $N=$ odd,
while all the components
are nonzero for $N=$ even. This implies, again, that the phase
structure is different, depending on whether $N=$ odd or even.
\par
%%%%%%%%
If $N=$ odd, it is possible for the Higgs VEV to take nonzero
values, keeping the cross term vanishing and the Higgs potential
minimizing, thanks to the zero in Eq.(\ref{configadj}).
Then, one of the $U(1)$'s is broken by
the Higgs VEV, so that the residual gauge symmetry is $U(1)^{N-2}$.
It is important to note that in the model there is only one phase
whose structure does not depend on the size of the extra
dimension. This is 
a new feature that has not
been observed in the phase structures obtained in the previous section.
If $N=$ even $(\geq 4)$, there appears no zero component in
Eq.(\ref{configadj}). One needs to study the effective potential
carefully in order to determine the vacuum structure of the model.
In case of the $SU(2)$ gauge group with
$N_F$ massless adjoint fermions,
we can perform fully analytic calculations,
and the phase structure
is found to be very similar to
the one obtained in the previous paper,
though the residual gauge symmetry in the Hosotani phase
is given by $U(1)$ in this case.
\par
%%%%%%%%%%%%
Let us next consider the adjoint Higgs instead of the fundamental Higgs.
Both $\vev{A_y}$ and $\vev{\Phi}$ belong to the adjoint representation
under the $SU(N)$ gauge group. The cross term corresponding to the third
term in the effective potential (\ref{effpotu}) is replaced by
%%%%%%%%%
\begin{equation}
g^2{\rm tr}\left([\vev{A_y},~\vev{\Phi}]^2\right),
\end{equation}
which is positive semidefinite. The diagonal
form of $\vev{\Phi}$ makes the term vanish to minimize
the effective potential. Then, the effective potential is
divided into two parts written in terms of $\theta_i$ or $v$ alone.
As a result, the minimization can be carried out separately with
respect to the order parameters, so that
the phase structure
does not depend on the size of $S^1$. The residual gauge
symmetry is determined by
the generators of $SU(N)$ which commute with
both $\vev{A_y}$ and $\vev{\Phi}$. Though we have already known that
the gauge symmetry is (un)broken to $(SU(N))$ $U(1)^{N-1}$ through the
Hosotani mechanism 
if the fermions belong
to the (fundamental) adjoint representation under $SU(N)$, the
actual residual gauge symmetry depends on the structure
of the Higgs potential.
\par
%%%%%%%%%%%
If we assume the same type of
the Higgs potential
as Eq.(\ref{potential}), for example,
%there is 
there are
generally flat directions
parametrized like $\vbar_1^2 +\cdots +\vbar_{N-1}^2
+(\vbar_1 +\cdots + \vbar_{N-1})^2=\mbar^2/\lambda$
in the effective potential. The residual gauge symmetry
%is not determined 
is not uniquely determined
in this case\footnote{This is the case
within the approximation we have made.}.
For the fermions
belonging to
the fundamental representation under $SU(N)$, depending on the
form of the Higgs VEV, the $SU(N)$ gauge symmetry is broken to
%some of its Cartan subgroup.
its subgroup.
On the other hand, 
for the fermions
in the adjoint representation, the residual gauge symmetry
is $U(1)^{N-1}$, irrespective of
%the flat direction.
the flat directions.
%%%%%%%%%%%%%%%%%%%%%%%%%%%%%%%%%%%%%%%%%
%%%%%%%%%%%%%%%%%%%%%%%%%%%%%%%%%%%%%%%%%
\section{Conclusions and Discussions}
%%%%%%%%%%%%%%%%%%%%%%%%%%%%%%%%%%%%%%%%%
%%%%%%%%%%%%%%%%%%%%%%%%%%%%%%%%%%%%%%%%%
We have study the phase structure of the $SU(N)$ gauge-Higgs 
models with $N_F$ massless fermions
on the space-time $M^3\otimes S^1$.
There are two kinds of
the order parameters
for gauge symmetry breaking
in the models.
One is the vacuum expectation value of the Higgs field
$\vev{\Phi}$ (the Higgs mechanism)
and the other
is the vacuum expectation value of the
component gauge field for the $S^1$ direction
$\vev{A_y}$ (the Hosotani mechanism).
The former works at the tree level, while the latter
is effective at the quantum level and sensitive to the size
of $S^1$. There is also the interaction
between $\vev{A_y}$ and $\vev{\Phi}$, which depends on the size as well.
Thus, the dominant contribution to the effective potential comes
from the different physical origins,
depending on the size of the extra dimension.
Therefore, the phase structure depends on the size
(in addition to 
the parameters of the models) in general. This is
expected to be a general feature in gauge-Higgs models on the space-time.
\par
%%%%%%%%%%%%%
We have computed the effective potential for the two kinds of
the order parameters in a one-loop approximation.
In the calculation we have assumed that the number of the massless
fermions is large enough, so that we have neglected the one-loop
contributions from the gauge and
Higgs fields to the effective potential. Then, we have obtained the
effective 
potential given by Eq.(\ref{effpotu}). It turns out that
the existence of the cross term in the potential, which comes
from the interaction between $\vev{A_y}$ and $\vev{\Phi}$ in
%$D_{\hmu}\Phi$, is crucial
$D_{y}\Phi$, plays a crucial role
to determine the phase structure of the model.
\par
%%%%%%%%%%%
We have first considered the case that both the fermion and Higgs fields
belong to the fundamental representation under $SU(N)$.
The model possesses
three phases called Hosotani, Higgs
and coexisting phases for $N=$ odd, while for $N=$ even the model has only
two 
phases, Hosotani and coexisting phases. The Higgs phase does not exist
for finite sizes of $S^1$ and $N=$ even.
The phase structure
depends on both the size of the extra dimension and
the parameter of the model.
We have 
obtained
the phase structure depicted
in Fig. 2 (3)
for $N=$ odd (even). It should be noted that, contrary to the usual
case, the Hosotani mechanism can play a role of
the restoration of gauge symmetry in the
model.
\par
%%%%%%%%
We have next considered the case that the representation
of the fermions
is changed into the adjoint representation under $SU(N)$.
The $SU(N)$ gauge symmetry is maximally
broken to $U(1)^{N-1}$ through the Hosotani mechanism. If $N=$ odd, the
Higgs
field can acquire 
the nonvanishing vacuum expectation value,
keeping the cross term
vanishing. Then, one of the $U(1)$'s is further broken by the
Higgs VEV, so that the residual gauge symmetry is $U(1)^{N-2}$.
There is only one phase in the model, which does not depend on
the size of the extra dimension. On the other
hand, if $N=$ even $(\geq 4)$, due to the nonexistence of the zero component
in the $\vev{A_y}$ unlike the case $N=$ odd,
one has to study the effective potential carefully in order to
investigate the phase structure of the model. The phase structure
for $SU(2)$, however,
can be fully studied analytically and
is similar to the one obtained
in the previous paper \cite{previous}.
The residual gauge symmetry in the Hosotani phase is given
by $U(1)$ in this case.
\par
%%%%%%%%%%%%%
We have also considered the case that the Higgs field belongs to the
adjoint representation under $SU(N)$. Both $\vev{A_y}$
and $\vev{\Phi}$ belong to the
adjoint representation, and they cannot be diagonalized
simultaneously, in general.
The cross term, however, requires the
diagonal form of $\vev{\Phi}$ in order for the effective potential to be
minimized. Then, the effective potential is separated into two parts with
respect to the order parameters
in our approximation, and
the minimization
of the potential is carried out separately.
This implies
that the phase structure of the model
does not depend
on the size of the extra dimension and there is only one phase in the
model. The residual gauge symmetry
in the phase is generated by
the generators of $SU(N)$
commuting with
both $\vev{A_y}$ and $\vev{\Phi}$, and it depends on the
detailed structure of the Higgs potential.
\par
%%%%%%%%%%%%%%
Our models have been studied on the space-time $M^3\otimes S^1$.
One may wonder what will happen if we consider
models
on $M^4\otimes S^1$, or more
generally, $M^{D-1}\otimes S^1$.
Qualitative features
such as the existence of
the several phases
and their structures
with respect to the scale
will not change even if we go to
the higher dimensions. The phase structure comes
from the fact that each term in the effective potential (\ref{effpotu}) has
the different dependence on the size of the extra dimension.
In other words, each term has its own physical origin, which is
different each other
and exists even in the higher dimensions.
If we start with the space-time $M^{D-1}\otimes S^1$, the fermion
one-loop correction is given by
%%%%%%%
\begin{equation}
{{2^{[{D\over 2}]}
N_F \Gamma(\frac{D}{2})}
\over{\pi^{{D\over 2}}L^D}}
\sum_{i=1}^{N}\sum_{n=1}^{\infty}{1\over n^D}\cos(n\theta_i).
\end{equation}
The scale of 
the term
is governed by the factor $1/L^D$ in place of $1/L^4$ in $D$ dimensions.
The minimum 
of this function is given by the same configuration
as Eqs.(\ref{hosotani})
or
(\ref{hosotanid}) 
in Sec. 2. This means that the global minimum
for the correction does not depend on the total dimension.
This is because the Hosotani mechanism
is controlled by the infrared physics like the Casimir
effect. In fact, the one-loop potential is governed only by the
light modes in the Kaluza-Klein ones, so that they mainly contribute
to determine the dynamics. On the contrary, the heavy modes
suppresses the effective potential more as the dimension becomes higher.
The cross term, which is crucial for the phase structure, also exists
even in the higher dimensions as the same way.
Therefore, we expect
that the qualitative features found in this paper
do not change even if we start with the higher dimensions.
\par
%%%%%%%%%%%%%%% 
In computing the effective potential, we have neglected the one-loop
corrections to $\vev{A_y}$ from the gauge and Higgs fields.
We have assumed that the number of
massless 
fermions
is large enough, so that these contributions are suppressed.
One needs to 
take account of
these contributions to the effective potential for small $N_F$ in
order to understand the whole
vacuum structure of the model. Namely, it is expected that
the phase structure for small radius of $S^1$ is
more involved because the ignored terms
start to come into play in the effective potential.
\par
%%%%%%%%%%
We have also ignored the one-loop corrections to the Higgs potential
from the gauge and Higgs fields, such as
$c\frac{g^2}{L^2}\Phi^{\dagger}\Phi$ and
$c^{\prime}\frac{\lambda}{L^2}\Phi^{\dagger}\Phi$,
by assuming that the couplings $g$ and $\lambda$ are sufficiently
small.
Those mass corrections are irrelevant to the model considered in Sec.2,
but they could cause gauge symmetry restoration at very small scales
for models with nonvanishing Higgs VEV, like the models with
one phase found in Sec.3.
Mass corrections to Higgs potentials at finite temperatures or
finite scales of extra dimensions have been investigated in
many literature \cite{O(N) model, finitetemperature, finiteradius}
and their effects on gauge symmetry breaking/restoration are
well understood.
Since the subject is not our main concerns, we will not discuss it
any more.
\par
%%%%%%%%%
There are several directions to extend our studies.
It is interesting to investigate
how gauge symmetry breaking patterns can be rich in
the phase diagram by introducing matter fields belonging to
various representations of gauge groups\footnote{
The twisted 
boundary condition of matter for the $S^1$ direction also affects the
phase structure \cite{SUSY, translation, O(N) model}.}. 
This study has the relation with
the new approach to the gauge hierarchy problem we
proposed in the previous paper. We have considered the massless
fermions
through the analyses. In connection with the suppression of
the effective potential by the large fermion number, a massive fermion
also modifies the size of the fermion one-loop correction
like $\e^{-mL}/L^4$ for $L>m^{-1}$. It is interesting to see how
massive fermions affect
the phase structure.
We should finally stress that if the Standard Model were embedded
in a higher dimensional theory with a multiply connected space,
our studies would have physical importance because the theory
is just in a class of the gauge-Higgs system on multiply
connected spaces.
It would be of importance to investigate the phase structure and
clarify its physical consequences at low energies.
Those will be reported elsewhere.
%%%%%%%%%%%%%%%%%%%%%%%%%%%%%%%
\vskip 2cm
\begin{center}
{\bf Acknowledgements}
\end{center}  
K.T would like to thank the Dublin Institute for Advanced Study
for warm hospitality, where part of this work was done.
This work was supported in part by a JSPS Research Fellowship
for Young Scientists (H.H).
%\vskip 2cm
\newpage
%%%%%%%%%%%%%%%%%%%%%%%%%%%%%%%%%%%%%%%%%%%
%%%%%%%%%% HERE COMES APPENDIX %%%%%%%%%%%%
%%%%%%%%%%%%%%%%%%%%%%%%%%%%%%%%%%%%%%%%%%%
\begin{center}
{\Large\bf Appendix}
\end{center}
%%%%%%%%%%%%%%%%
\begin{flushleft}
(A)~~{\it Parametrization of
the vacuum expectation value for the Higgs field}
\end{flushleft}
%%%%%%%%%%%%%%%%
We shall show that the parameterization (\ref{vev}) of the
vacuum expectation value for the Higgs field can minimize
the effective potential (\ref{effpotu}).
\par
%%%%%%%%%
The classical part of the effective potential for
$\vev{A_y}$ and $\vev{\Phi}$ with every position being filled
is given by
%%%%%%%%%%%
\begin{eqnarray}
V_{cl}&=&-\half m^2\sum_{i=1}^{N}|v_i|^2
          + {\lambda\over 8}\left(\sum_{i=1}^{N} |v_i|^2\right)^2
          +\frac{1}{2L^2}\sum_{i=1}^N\hat{\theta}_i^2 |v_i|^2 ,
\label{classical}
\end{eqnarray}
where $\hat{\theta}_i = \theta_i$ mod $2\pi$ with
$|\hat{\theta}_i| \le \pi$ for $i=1,\cdots, N$.
Without loss of generality, we can assume that
$\abs{\hat{\theta}_1}\leq\abs{\hat{\theta}_2}\leq\cdots\leq
\abs{\hat{\theta}_{N}}$.
Then, it is convenient to rewrite $V_{cl}$ into the form
%%%%%%%%%
\begin{eqnarray}
V_{cl} &=& V_1 +V_2 ,
\end{eqnarray}
where
%%%%%%%%%
\begin{eqnarray}
V_{1}&=&-\frac{1}{2L^2}( m^2 L^2 - \hat{\theta}_{1}^{2})
         \sum_{i=1}^{N}|v_i|^2
          + {\lambda\over 8}\left(\sum_{i=1}^{N} |v_i|^2\right)^2 ,
           \\
V_{2}&=& \frac{1}{2L^2}
        \sum_{i=2}^{N}(\hat{\theta}_i^2 - \hat{\theta}_{1}^{2})|v_i|^2 .
\label{V1V2}
\end{eqnarray}
Note that $V_1$ depends only on $\sum_{i=1}^{N}|v_i|^2 $ 
(and $\hat{\theta}_i$)
and $V_2$ is positive semidefinite.
\par
%%%%%%%%%%
Let us now consider a minimization problem of $V_{cl}$ for fixed
$\theta_i$ ($i=1,\cdots, N$).
Suppose that the minimum of $V_1$ for fixed $\theta_i$ is realized
by $\sum_{i=1}^{N} |v_i|^2 = v^2$ for a real constant $v$.
Then, it is easy to see that the configuration
$\vev{\Phi} = \frac{1}{\sqrt{2}}(v_1, 0,\cdots, 0)^T$ with
$|v_1|^2 = v^2$ gives the minimum of $V_{cl}$ for fixed $\theta_i$,
because the configuration realizes the minimum values of
$V_1$ and $V_2$ simultaneously, so that it must be a configuration
which minimizes $V_{cl}$.
By using a $U(1)$ symmetry to make $v_1$ real,
we arrive at the expression (\ref{vev}).
Since the incorporation of quantum corrections to $\vev{A_y}$
does not alter the above discussion, we have proved the
parameterization (\ref{vev}) in the text.
\par
%%%%%%%%%%%%%%%%%%%%%%
\begin{flushleft}
(B)~~{\it Expressions and results}
\end{flushleft}
%%%%%%%%%%%%%%%%%%%%%%
We shall derive some expressions and results used in the text.
%%%%%%%%%%%%%%
\begin{flushleft}
(B)-1~~{\it Hosotani phase and its stability}
\end{flushleft}
%%%%%%%%%%%%%%%%%
The Hosotani phase is obtained by solving the
equation (\ref{eqhosotani}) and the Higgs
VEV is given by Eq.(\ref{hvevu}) in the text.
We work on the space-time $M^3\otimes S^1$, so that there is a
formula,
%%%%%%%
\begin{equation}
\sum_{n=1}^{\infty}{1\over n^4}\cos(nx)={{-1}\over {48}}x^2(x-2\pi)^2
+{{\pi^4}\over {90}}\qquad \mbox{for} \ 0\le x \le 2\pi
\label{formulad}
\end{equation}
from which we have 
%%%%%%%%%%%
\begin{eqnarray}
\sum_{n=1}^{\infty}{1\over n^3}\sin(nx)&=&{1\over {12}}x(x-\pi)(x-2\pi),
\label{formulau}\\
%%%%%%%
\sum_{n=1}^{\infty}{1\over{n^2}}\cos(nx)&=&{1\over 4}x(x-2\pi)+{\pi^2\over
6}.
\label{formulat}
\end{eqnarray}
Note that the minimum of the function (\ref{formulad}) is
located at $x=\pi$.
\par
%%%%%%%%%%
To make our analysis simple, let us assume that
$\theta_i (i=1,\cdots, N-1)$ lies in the range of $0\le\theta_i<2\pi$.
This can be done without loss of generality.
Applying the formula (\ref{formulau}) to Eq.(\ref{eqhosotani}),
we obtain 
%%%%%%%
\begin{eqnarray}
&&\biggl(\theta_k +\sum_{i=1}^{N-1}\theta_i-2\pi q\biggr)
  \biggl(\theta_k^2+\Bigl(\sum_{i=1}^{N-1}\theta_i
   - 2\pi(q-1)\Bigr)^2 - \theta_k\Bigl(\sum_{i=1}^{N-1}\theta_i
   -2\pi(q-1)\Bigr)\nonumber\\
&&\qquad   
   -\pi \theta_k - \pi \Bigl(\sum_{i=1}^{N-1}\theta_i
   - 2\pi(q-1)\Bigr)\biggr)
     =0 \qquad \mbox{for}\  k=1,2,\cdots, N-1,
\label{extuno}
\end{eqnarray}
where the integer $q$ is defined by the requirement
%%%%%%%%%
\begin{eqnarray}
0 \le \sum_{i=1}^{N-1} \theta_i - 2\pi (q-1) < 2\pi .
\end{eqnarray}
Let us first study the case given by
%%%%%%%%%
\begin{eqnarray}
\theta_k + \sum_{i=1}^{N-1} \theta_i = 2\pi q\qquad
    \mbox{for}\ k=1,2,\cdots, N-1.
    \label{extdue}
\end{eqnarray}
The solution to the equation is obtained as
%%%%%%%%%
\begin{eqnarray}
\theta \equiv \theta_k =\frac{2\pi q}{N}\ \ (k=1,\cdots, N-1),
\quad q=0,1,\cdots, N-1.
\end{eqnarray}
Knowing that the effective potential is, now, recast
in
%%%%%%%%
\begin{equation}
\Vb={{AN}\over{\pi^2}}\sum_{n=1}^{\infty}{1\over n^4}
\cos(n{{2\pi q}\over{N}}),
\end{equation}
we find that the potential is minimized at $q=\frac{N-1}{2}$
for $N=$ odd and at $q=\frac{N}{2}$ for $N=$ even.
Thus, we have
%%%%%%%
\begin{equation}
(\vbar, \theta)=\left\{\begin{array}{ll}
(0,~~{{N-1}\over N}\pi),& N={\rm odd},\\[0.4cm]
(0,~~\pi),& N={\rm even},\end{array}\right.
\label{solap}
\end{equation}
which give the solutions  (\ref{hosotani}) and (\ref{hosotanid})
corresponding to the Hosotani phase in the text.
\par
%%%%%
Let us next discuss
%stability of the solution
the stability of the above solution
against small fluctuations.
The stability is guaranteed
if all the eigenvalues of the
Hessian is positive definite. The Hessian is given by
the second derivative of the effective
potential with respect to the order parameters,
%%%%%%%%%%%
\begin{equation}
H \equiv 
\left(\begin{array}{cccc}
{{\del^2\Vb}\over{\del\vbar^2}} &{{\del^2\Vb}\over{\del\vbar\del\theta_1}} &
\cdots & {{\del^2\Vb}\over{\del\vbar\del\theta_{N-1}}} \\
{{\del^2\Vb}\over{\del\vbar\del\theta_1}}
&{{\del^2\Vb}\over{\del\theta_1^2}} &
\cdots & {{\del^2\Vb}\over{\del\theta_1\del\theta_{N-1}}} \\
\vdots & \vdots & \ddots & \vdots \\
{{\del^2\Vb}\over{\del\vbar\del\theta_{N-1}}} &
{{\del^2\Vb}\over{\del\theta_1\del\theta_{N-1}}} &
\cdots & {{\del^2\Vb}\over{\del\theta^2_{N-1}}}
\end{array}\right).
\end{equation}
The matrix $H$ evaluated at the solution becomes
%%%%%%%%%%%%%%%%%%
\begin{equation}
H=\left(\begin{array}{ccccc}
F & 0 &\cdots &\cdots & 0\\
0                          &2C & C     &\cdots &C \\
\vdots                     &C  &\ddots & &C \\
\vdots                     &\vdots & & \ddots&\vdots\\
0                          &C      &\cdots&\cdots&2C\\
\end{array}\right),
\end{equation}
where we have defined
%%%%%%%%
\begin{equation}
F\equiv -\mbar^2+\left({{2\pi q}\over{N}}\right)^2,~~C\equiv {A\over\pi^2}
\sum_{n=1}^{\infty}{{-1}\over n^2}\cos\left(n{{2\pi q}\over N}\right)
\end{equation}
with $q={{N-1}\over 2}$ or ${N\over 2}$.
The eigenvalues of the matrix $H$ are found to be
$F$, $C$ ($(N-2)$ degeneracy) and $NC$, and hence
all the eigenvalues
of $H$ for the given values of $q$ is positive as long
as $0 < \mbar < {{2\pi q}\over N}$.
This means that each solution in Eq.(\ref{solap}) is stable for the
scale region given by
%%%%%%%
\begin{equation}
\begin{array}{lll}
0< \mbar < {{N-1}\over N}\pi \quad
& {\rm for }\quad & N={\rm odd},\\[0.3cm]
0< \mbar < \pi \quad & {\rm for }\quad & N={\rm even}.
\end{array}
\end{equation}
We have obtained Eqs.(\ref{regionu}) and (\ref{regionq}) in the text.
\par
%%%%%%%%
Let us next consider the other solutions to Eq.(\ref{extuno}),
%%%%%%%%
\begin{eqnarray}
&&\biggl(\theta_k^2+\Bigl(\sum_{i=1}^{N-1}\theta_i
   - 2\pi(q-1)\Bigr)^2 - \theta_k\Bigl(\sum_{i=1}^{N-1}\theta_i
   -2\pi(q-1)\Bigr)\nonumber\\
&&\quad   -\pi \theta_k - \pi \Bigl(\sum_{i=1}^{N-1}\theta_i
   - 2\pi(q-1)\Bigr)\biggr)
     =0 
\label{altrosol}
\end{eqnarray}
for $k=1,\cdots, N-1$.
Any solutions satisfying Eq.(\ref{altrosol}) give a negative
diagonal component $\frac{\partial^2 \bar{V}}{\partial \theta_k^2}$
in $H$.
It is not difficult to show that
%%%%%%%%%%%
\begin{eqnarray}
{{\del^2\Vb}\over{\del\theta_k^2}}
  &=&
   {A\over\pi^2}\sum_{n=1}^{\infty}
     {{-1}\over{n^2}}
       \biggl(\cos(n\theta_k)+\cos(n\sum_{i=1}^{N-1}\theta_i)\biggr)
       \nonumber\\
  &=& 
   -{{A}\over{12\pi^2}}
     \biggl(\theta_k +\sum_{i=1}^{N-1}\theta_i -2\pi q \biggr)^2 <0,
\end{eqnarray}
where we have used the formula (\ref{formulat}) and Eq.(\ref{altrosol}).
This implies that any solutions satisfying Eq.(\ref{altrosol}) are
not stable against small fluctuations, so that we exclude such
solutions from our discussions hereafter.
%%%%%%%%%%%%%
\begin{flushleft}
(B)-2~~{\it Higgs and coexisting phases and their stabilities}
\end{flushleft}
%%%%%%%%%%%%%
Let us next consider the case given by Eq.(\ref{hvevd}),
%%%%%%%%
\begin{equation}
-\mbar^2+{\lambda\over 2}\vbar^2 + \hat{\theta}_1^2=0.
\label{altrovev}
\end{equation}
In this case, the equations we solve are given by Eqs.(\ref{eqcoexistu})
and (\ref{eqcoexistd}). Applying the formula (\ref{formulau})
to Eq.(\ref{eqcoexistd}), we obtain the same equation
as Eq.(\ref{extdue}), in which the case of $k=1$ is excluded
in the present case.
The equations obtained imply that
%%%%%%%
\begin{equation}
\theta_2=\theta_3=\cdots = \theta_{N-1}\equiv \btheta.
\label{samecond}
\end{equation} 
As the result, 
the equation (\ref{eqcoexistd}) finally yields the relation,
%%%%%%%%%
\begin{equation}
\theta_1+(N-1)\btheta=2\pi l
\label{relation}
\end{equation} 
for some integer $l$.
Since it is enough to consider the region $0\leq \bar{\theta} \leq \pi$
and we have required the sequence
$\abs{\hat{\theta}_1}\leq\abs{\hat{\theta}_2}\leq\cdots\leq
\abs{\hat{\theta}_{N}}$,
our solutions also have
to satisfy the constraint
$\abs{\hat{\theta}_1}\leq \abs{\btheta}$
in addition to the relation (\ref{relation}).
Among possible solutions satisfying those, one needs the solution
that minimizes the effective potential,
which is now recast in
%%%%%%%
\begin{equation}
\Vb=-\half \mbar^2\vbar^2 + {\lambda\over 8}\vbar^4
    + \half \hat{\theta}_1^2 \vbar^2
     +{A\over \pi^2}\sum_{n=1}^{\infty}{1\over n^4}
\left(\cos(n\theta_1) + (N-1)\cos(n\btheta))\right),
\label{effpotential1}
\end{equation} 
where we have used 
Eqs.(\ref{samecond}) and (\ref{relation}).
%%%%%%%%%%%
The integer $l$ must be determined in such a way that the potential
energy is minimized. For general $l$, Eq.(\ref{relation}) and 
other constraints restrict
allowed regions of $\theta_1$ and $\bar{\theta}$.
It is not difficult to see that 
the minimum of the effective potential (\ref{effpotential1}) can 
be realized when $l={{N-1}\over 2}$ for $N=$odd 
or $l={N\over 2}$ for $N=$ even and
%%%%%%%%%%
\begin{eqnarray}
{{N-1}\over N}\pi \leq \btheta \leq \pi, &&
\quad 0\leq \theta_1 \leq {{N-1}\over N}\pi,\quad
\mbox{for}\ N=\mbox{odd},
\label{desiu}\\
{{N-2}\over{N-1}}\pi \leq \btheta \leq \pi,&&
\quad \pi \leq \theta_1 \leq 2\pi, \hspace{1.3cm}
\mbox{for}\ N=\mbox{even}.
\label{desid}
\end{eqnarray}
Thus, we have obtained Eqs.(\ref{restrictodd})
and (\ref{restricteven}) in the text.
\par
%%%%%%%%%%%%%%
Now, the effective potential is rewritten, depending on whether $N$
is even or odd, in terms of $\vbar$ and $\theta_1$ alone, as
%%%%%%%%%%%%%
\begin{eqnarray}
\Vb_{N=odd}&=&-\half \mbar^2\vbar^2 + {\lambda\over 8}\vbar^4 +
\half \hat{\theta}_1^2 \vbar^2\nonumber\\
&+&{A\over \pi^2}\sum_{n=1}^{\infty}{1\over n^4}
\biggl(\cos(n\theta_1)
+ (N-1)\cos\Bigl(n\bigl(\pi +{{\theta_1}\over{N-1}}\bigr)\Bigr)
\biggr),\label{effapu}\\
%%%%%%%%
\Vb_{N=even}&=&-\half \mbar^2\vbar^2 + {\lambda\over 8}\vbar^4 +
\half {\hat{\theta}}_1^2 \vbar^2\nonumber\\
&+&{A\over \pi^2}\sum_{n=1}^{\infty}{1\over n^4}
\biggl(\cos(n\theta_1) + (N-1)\cos\Bigl(n\bigl({{\theta_1}\over{N-1}}+
{{N-2}\over{N-1}}\pi\bigr)\Bigr)
\biggr).\label{effapd}
\end{eqnarray}
It follows from Eqs.(\ref{desiu}) and (\ref{desid})  that
the relation between $\hat{\theta}_1$ and $\theta_1$ is given
by $\hat{\theta}_1=\theta_1$ for $N=$ odd and
$\hat{\theta}_1=\theta_1 -2\pi$ for  $N=$ even.
Our remaining task
is to solve the equation (\ref{eqcoexistu}) under the
relation (\ref{relation}) with
$q=\frac{N-1}{2}$ $(\frac{N}{2})$ for $N=$ odd (even)
or, equivalently, to solve
the equation
from the first derivative of the potential
(\ref{effapu}) or (\ref{effapd}) with respect to $\theta_1$
with Eq.(\ref{altrovev}).
%%%%%%%%%%%%%%%
%%%%%%%%%%%%%%
\begin{flushleft}
(B)-3~~{\it $N=$odd}
\end{flushleft}
%%%%%%%%%%%%%%%
%%%%%%%%%%%%%%
As explained above, the equation we solve becomes
%%%%%%%%%%%%
\begin{equation}
\theta_1\left(
{2\over \lambda}(\mbar^2-\theta_1^2)-{A\over{12\pi^2}}
\Bigl({{N(N^2-3N+3)}\over{(N-1)^3}}\theta_1^2
-3\pi\theta_1+{{2N-3}\over {N-1}}\pi^2\Bigr)
\right)=0,
\label{cohiggsu}
\end{equation}
which reads 
%%%%%%%
\begin{eqnarray}
&&\theta_1=0,\label{higgssou}\\
&\mbox{or}&\nonumber\\
&&\left(1+{t\over{24\pi^2}}{{N(N^2-3N+3)}\over{(N-1)^3}}\right)\theta_1^2
-{t\over{8\pi}}\theta_1-\mbar^2 +{t\over{24}}{{2N-3}\over{N-1}}=0.
\label{kyouzon}
\end{eqnarray}
Here we have introduced $t\equiv \lambda A(=4\lambda N_F)$.
\par
%%%%%%%%%%%%%
Let us first study the case $\theta_1=0$. The relation (\ref{relation})
with $l={{N-1}\over 2}$
yields $\theta_2=\theta_3=\cdots=\theta_{N-1}=\pi$
and $\theta_N=-(N-2)\pi$. And $\vbar=\sqrt{2/\lambda}~\mbar$
from Eq.(\ref{altrovev}). Thus, we have obtained the
type II solution corresponding to the Higgs phase (\ref{higgs})
in the text. The stability of the type II solution is
studied by the eigenvalues of the
matrix $H$ evaluated at the solution. It is given by
%%%%%%%%%
\begin{equation}
H=
\left(\begin{array}{ccccc}
2\mbar^2 & 0 &\cdots &\cdots & 0\\
0        &G & B &\cdots &B \\
\vdots   &B                           &2B & &B \\
\vdots   &\vdots & & \ddots&\vdots\\
0        &B      &\cdots&\cdots&2B\\
\end{array}\right),
\end{equation}
where we have defined $B\equiv {A\over{12}}$ and $G\equiv
{2\over \lambda}\mbar^2-B$.
The eigenvalues of the matrix
are found to be 
$2\bar{m}^2$,
${A\over{12}}$ ($(N-3)$ degeneracy) and $x_{\pm}$, where
%%%%%%%%%
\begin{equation}
x_{\pm}\equiv\half\Biggl(G+(N-1)B
\pm \sqrt{(G+(N-1)B)^2-4B((N-1)G-(N-2)B)}\Biggr).
\end{equation}
The condition that the eigenvalues $x_{\pm}$ are positive is given by
%%%%%%%%
\begin{equation}
\mbar^2 > {{2N-3}\over{N-1}}{{\lambda A}\over{24}}
={{2N-3}\over{N-1}}{t\over{24}}\equiv \mbar_3^2.
\end{equation}
Thus, we have obtained Eq.(\ref{regiond}) in the text.
\par
%%%%%%%%%%%%%%%%
Let us next study 
the solutions
given by Eq.(\ref{kyouzon}),
{\it i.e.}
%%%%%%%
\begin{equation}
\theta_1^{\pm}(\mbar)=
{1\over {2(1+{{t}\over{24\pi^2}}C_0)}}\left(
{{t\over{8\pi}}\pm {{N^2-3}\over{24(N-1)^2}}{t\over \pi}
\sqrt{S(\mbar)}}\right),
\label{cosol}
\end{equation}
%%%%%%% 
where
\begin{equation}
S(\mbar)\equiv 1-C_1\left({\pi^2\over t}\right)+
{1\over t^2}\left(C_2\pi^2 +C_3 t \right)\mbar^2
\end{equation}
with $C_i(i=0,1,2,3)$ being defined by
%%%%%%%
\begin{eqnarray}
C_0 &\equiv & {{N(N^2-3N+3)}\over {(N-1)^3}},~~
C_1\equiv {{96(2N-3)(N-1)^3}\over{(N^2-3)^2}},\\
C_2 &\equiv & {{
2304(N-1)^4}\over{(N^2-3)^2}},~~
C_3\equiv {{96 N(N-1)(N^2-3N+3)}\over{(N^2-3)^2}}.
\end{eqnarray}
Let us study the stability of
the solutions.
To this end, we note
that the order parameters in this case are reduced to
two, that is, $\vbar$ and $\theta_1$. Then, the matrix $H$ becomes
the $2\times 2$ matrix, 
%%%%%%%%%%%
\begin{equation}
H \equiv 
\left(\begin{array}{cc}
{{\del^2\Vb}\over{\del\vbar^2}} &
{{\del^2\Vb}\over{\del\vbar\del\theta_1}}\\
{{\del^2\Vb}\over{\del\vbar\del\theta_1}}
&{{\del^2\Vb}\over{\del\theta_1^2}}\\
\end{array}\right),
\label{hessian2}
\end{equation}
where each component evaluated at
the solutions
is given by
%%%%%%%%%%
\begin{eqnarray}
{{\del^2\Vb}\over{\del\vbar^2}}&=&\lambda\vbar^2,\qquad
{{\del^2\Vb}\over{\del\vbar\del\theta_1}}=2\theta_1^{\pm}\vbar\\
{{\del^2\Vb}\over{\del\theta_1^2}}&=&
\vbar^2 -{A\over{\pi^2}}\sum_{n=1}^{\infty}{1\over n^2}
\left(\cos(n\theta_1^{\pm})+
{1\over{N-1}}\cos(\pi+{\theta_1^{\pm}\over{N-1}})\right)
\nonumber\\
&=&{A\over{12\pi^2}}
\left(-{{2N(N^2-3N+3)}\over{(N-1)^3}}(\theta_1^{\pm})^2+3\pi\theta_1^{\pm}
\right),
\end{eqnarray}
where we have used the formula (\ref{formulat}).
Then, the determinant of $H$ is calculated as
%%%%%%%%
\begin{equation}
{\rm det}~H= \mp\vbar^2(\mbar)\theta_1^{\pm}(\mbar)
\left(
{{t(N^2-3)}\over{12\pi (N-1)^2}}\sqrt{S(\mbar)}\right).
\end{equation}
Since $\theta_1$ is larger than zero, the solution $\theta_1^+$ gives
a negative determinant of $H$, so that
$\theta_1^+(\mbar)$ is unstable and is
excluded from our 
discussions.
Hence, we have obtained the type III
solution $\theta_1^-(\mbar)$ with $\vbar =\sqrt{{2\over\lambda}
(\mbar^2-(\theta_1^-)^2)}$ corresponding to the
coexisting phase (\ref{coexist}) in the text.
\par
%%%%%%%%%%%%%  
The solution $\theta_1^-(\mbar)$ must satisfy the
reality condition $(\theta_1^-(\mbar))^*=\theta_1^-(\mbar)$ and
$ 0 \leq \theta_1^-(\mbar) \leq {{N-1}\over N}\pi$, as shown in
Eq.(\ref{desiu}). The reality condition is furnished if
$t\ge C_1 \pi^2$ or if
%%%%%%%%
\begin{equation}
\mbar\geq \left({{C_1\pi^2 t -t^2}\over{C_2\pi^2+C_3 t}}\right)^{\half}
\equiv \mbar_1
\qquad \mbox{for}\ t<C_1 \pi^2 .
\label{m_1}
\end{equation}
The condition
$0 \leq \theta_1^-(\mbar)$ yields that
%%%%%%%
\begin{equation}
\mbar \leq \left({{2N-3}\over{N-1}}{t\over{24}}\right)^{\half}\equiv
\mbar_3, 
\end{equation}
while the condition $\theta_1^-(\mbar) \leq {{N-1}\over N}\pi$ requires that
%%%%%%%%
\begin{equation}
\mbar \geq {{N-1}\over{N}}\pi \equiv \mbar_2
\quad {\rm for}\quad
t\geq 48\pi^2{{(N-1)^3}\over{N(N^2-3)}}.
\end{equation}
The latter condition is always satisfied
for $t< 48\pi^2{{(N-1)^3}\over{N(N^2-3)}}$.
\par
%%%%%%%%%%%
The relative magnitude in
the scales
$\mbar_i$ $(i=1,2,3)$ is important to
understand the allowed region of the coexisting phase. It is easy to
show that $\mbar_1 < \mbar_3$ is
always satisfied, irrespective of the values of $N$ and $t$, while the
relative magnitude of
$\bar{m}_2$ and $\bar{m}_3$
depends on the parameter $t$,
%%%%%%%
\begin{equation}
\mbar_2 \leq (>) \mbar_3 \quad {\rm for}  \quad t\geq (<)
24\pi^2{{(N-1)^3}\over {N^2(2N-3)}}.
\end{equation}
The relation
$\mbar_1 \leq \mbar_2$ is always satisfied, where
the equality holds for $t=48\pi^2{{(N-1)^3}\over{N(N^2-3)}}$.
We have understood the scale
relations given in the parentheses
in Eqs.(\ref{cou}), (\ref{cod}) and (\ref{cot}).
%%%%%%%%%%%%%
In Fig.1, the curves of the critical scales $\bar{m}_i$ are depicted
in the $\bar{m}$-$t$ plane, and the relative magnitude of $\bar{m}_i$
will be understood clearly there.
Noting that $C_1\pi^2 > 48\pi^2\frac{(N-1)^3}{N(N^2-3)}$
and collecting the results obtained above,
we find that the allowed region of the coexisting phase 
%%%
%with the
%conditions $(\theta_1^{-}(\bar{m}))^{*} = \theta_1^{-}(\bar{m})$ and
%$0\le \theta_1^{-}(\bar{m})\le \frac{N-1}{N}\pi$ 
%%%%%%%
is given by
%%%%%%%%%
\begin{eqnarray}
\bar{m}_2 \le \bar{m} \le \bar{m}_3\qquad &\mbox{for}&\
   t > 48\pi^2\frac{(N-1)^3}{N(N^2-3)},\label{relativem1}\\
\bar{m}_1 \le \bar{m} \le \bar{m}_3\qquad &\mbox{for}&\
   t \le 48\pi^2\frac{(N-1)^3}{N(N^2-3)}.\label{relativem2}
\end{eqnarray}
It will be useful to evaluate the values of
$\theta_1^{-}(\bar{m})$ at the boundaries in
Eqs.(\ref{relativem1}) and (\ref{relativem2}).
One can show that
%%%%%%%%%%
\begin{eqnarray}
  \theta_1^{-}(\bar{m}_2) = \frac{N-1}{N}\pi\ \ \mbox{and}\ \
  \theta_1^{-}(\bar{m}_3) = 0 \quad
   &\mbox{for}&\ t\ge 48\pi^2\frac{(N-1)^3}{N(N^2-3)},\label{boundary1}\\
  \theta_1^{-}(\bar{m}_1) = \frac{t}
       {16\pi\left(1+\frac{t}{24\pi^2}C_0\right)}\ \ \mbox{and}\ \
  \theta_1^{-}(\bar{m}_3) = 0 \quad
  &\mbox{for}&\ t< 48\pi^2\frac{(N-1)^3}{N(N^2-3)}.\label{boundary2}
\end{eqnarray}
\par
%%%%%%%%%%
Let us study the behavior of the type III solution with respect to
the scale $\bar{m}$.
We first note that the solution $\theta_1^{-}(\bar{m})$ is
a monotonically decreasing function of $\bar{m}$.
%%%%%%%%%%
\begin{equation}
{{\del\theta_1^-(\mbar)}\over{\del\mbar^2}}=-
{{24\pi (N-1)^2}\over{t(N^2-3)}}{1\over{\sqrt{S(\mbar)}}}<0.
\end{equation}
On the other hand, $\bar{v}^2(\bar{m})$ is a monotonically
increasing function of $\bar{m}$
%%%%%%%%%%
\begin{equation}
{{\del\vbar^2}\over{\del\mbar^2}}
={2\over\lambda}\left(1+\theta_1^-(\mbar)
{{48(N-1)^2}\over{(N^2-3)\sqrt{S(\mbar)}}}{\pi\over t}\right)>0
\end{equation}
for the region (\ref{desiu}).
Since $\bar{v}^2(\bar{m})$ can be written as
%%%%%%%%%%
\begin{eqnarray}
\bar{v}^2(\bar{m}) 
  =
   \frac{N(N^2-3N+3)t}{12\pi^2(N-1)^3\lambda}
    \biggl(\theta_1^{-}(\bar{m}) - \frac{N-1}{N}\pi\biggr)
    \biggl(\theta_1^{-}(\bar{m}) - \frac{(N-1)(2N-3)}{N^2-3N+3}\pi\biggr),
\end{eqnarray}
$\bar{v}^2(\bar{m})$ is positive semidefinite for
$0\le \theta_1^{-}(\bar{m}) \le \frac{N-1}{N}\pi$,
as it should be.
One can also show that
%%%%%%%%%%%
\begin{eqnarray}
\bar{v}^2(\bar{m}_2) &=& 0\qquad
   \mbox{for}\ t\ge 48\pi^2\frac{(N-1)^3}{N(N^2-3)}, \\
\bar{v}^2(\bar{m}_3) &=& \frac{2}{\lambda}\bar{m}_3^2 .
\end{eqnarray}
It follows together with Eq.(\ref{boundary1}) and (\ref{boundary2})
that the coexisting phase is found to be continuously connected to the
Hosotani phase (the Higgs phase) at the boundary $\bar{m} = \bar{m}_2$
($\bar{m} = \bar{m}_3$) for
$t\ge 48\pi^2 \frac{(N-1)^3}{N(N^2-3)}$.
\par 
%%%%%%%%%%%%%%
We have studied the allowed region of the coexisting phase
with respect to the parameters $\bar{m}$ and $t$.
We have obtained that (i)
when $t > 48\pi^2{{(N-1)^3}\over{N(N^2-3)}}$,
the relative magnitude of the scales\footnote{
$\bar{m}_1$ is defined only for $t\le C_1\pi^2$.
} 
is given by $\mbar_1 < \mbar_2 < \mbar_3$, and
the coexisting phase exists between $\mbar_2$ and $\mbar_3$, (ii)
when $24\pi^2{{(N-1)^3}\over {N^2(2N-3)}}< t \leq
48\pi^2{{(N-1)^3}\over{N(N^2-3)}}$, the relative magnitude of
the scales
is given by $\mbar_1 \leq \mbar_2 < \mbar_3$, and the coexisting phase
lies between $\mbar_1$ and $\mbar_3$, (iii) when
$t \leq 24\pi^2{{(N-1)^3}\over {N^2(2N-3)}}$, we have
$\mbar_1 < \mbar_3 \leq \mbar_2$, and the coexisting phase is between
$\mbar_1$ and $\mbar_3$. We have arrived at the classification
used in the text, Eqs.(\ref{cou}), (\ref{cod}) and (\ref{cot}).
Fig.1 will help our understanding of the phase structure.
\par
%%%%%%%%%%%
Let us finally calculate the potential energy for each phase.
%%%%%%%%% 
\begin{eqnarray}
\Vb_{Hosotani}&=&A\pi^2
\left(-\frac{(N^2-1)^2}{48N^3}+{N\over{90}}\right),\\
%%%%%
\Vb_{Higgs}&=&-{\mbar^4\over{2\lambda}}+A\pi^2
\left(-{{(N-1)}\over{48}}+{{N}\over{90}}\right),\\
%%%%
\Vb_{coexisting}&=&-{{1}\over{2\lambda}}\left(\mbar^2-\theta_1^-(\mbar)^2
\right)^2
+{A\over\pi^2}\Biggl(-{1\over{48}}
{\theta_1^-(\mbar)}^2\Bigl(\theta_1^-(\mbar) - 2\pi\Bigr)^2 \nonumber \\
&&\ \ -{{N-1}\over{48}}\Bigl(\pi+{{\theta_1^-(\mbar)}\over{N-1}}\Bigr)^2
\Bigl({{\theta_1^-(\mbar)}\over{N-1}}-\pi\Bigr)^2
+{N\over{90}}\pi^4\Biggr),
\end{eqnarray}
%%%%%%%
where $\theta_1^-(\mbar)$ is given by Eq.(\ref{cosol}).
It is not difficult to show that the energy difference
$\Delta\Vb\equiv \Vb_{Hosotani}-\Vb_{coexisting}$ is a monotonically
increasing
function of $\bar{m}^2$,
%%%%%%%%%
\begin{equation}
{{\del}\over{\del\mbar^2}}\Delta \Vb(\mbar)
=\half \vbar(\mbar)^2 \geq 0,
\end{equation}
where we have used the equation (\ref{kyouzon}).
We also observe that
%%%%%%%%%
\begin{eqnarray}
\Vb_{Hosotani}-\Vb_{Higgs}
&=&{1\over{2\lambda}}
\left(\mbar^4 - 
{{(N-1)(N^2-N-1)}\over{N^3}}{{\pi^2}\over{24}}t\right)\nonumber\\
&\equiv & {1\over{2\lambda}}\left( \mbar^4 - (\mbar_5)^4\right),
\end{eqnarray}
which gives the critical scale given by Eq. (\ref{regiont}) in the text.
%%%%%%%%%%%%%%%%%%%%%%%%%
%%%%%%%%%% N=EVEN %%%%%%%%
%%%%%%%%%%%%%%%%%%%%%%%%%%
\begin{flushleft}
{\it (B)-4~~$N=$ even}
\end{flushleft}
Let us study the case $N=$ even. The equation we solve
is given, from Eq.(\ref{effapd}), by
%%%%%%%%%
\begin{eqnarray}
&-&{2\over\lambda}\left(\mbar^2-(2\pi-\theta_1)^2\right)(2\pi-\theta_1)
\nonumber\\
&+&{A\over\pi^2}
\sum_{n=1}^{\infty}{{-1}\over n^3}\left(\sin(n\theta_1)+
\sin(n({\theta_1\over{N-1}}+{{N-2}\over {N-1}}\pi))\right)=0,
\label{eqevenu}
\end{eqnarray}
where we have eliminated $\vbar^2$ by $\vbar^2={2\over\lambda}
(\mbar^2-(2\pi - \theta_1)^2)$.
Using the formula (\ref{formulau}), the above equation becomes
%%%%%%%%%%
\begin{equation}
\left(1+{{t\alpha}\over{24\pi^2}}\right)\theta_1^3
-\left({{t\alpha}\over{8\pi}} +6\pi\right)\theta_1^2
+\left({{t\beta}\over{24}}-\mbar^2+12\pi^2\right)\theta_1
+{{t\gamma}\over{24}}\pi +2\pi\mbar^2 -8\pi^3=0,
\label{eqevend}
\end{equation}
where 
%%%%%%%
\begin{equation}
\alpha \equiv  {{N(N^2-3N+3)}\over{(N-1)^3}},~
\beta \equiv {{N(2N^2-7N+8)}\over{(N-1)^3}},~
\gamma \equiv {{N(N-2)}\over{(N-1)^3}}.
\end{equation}
This is the equation (\ref{eqevendt}) in the text. Instead of solving
the equation (\ref{eqevend}) directly,
it turns out to be convenient to
study intersections of
two functions $F(\theta_1)$ and $G(\theta_1)$
followed from 
Eq.(\ref{eqevend}).
Here, $F(\theta_1)$ and $G(\theta_1)$ are
%%%%%%%%%%%%
\begin{eqnarray}
F(\theta_1)&\equiv &2(\mbar^2-(2\pi - \theta_1)^2)(2\pi -\theta_1),\\
G(\theta_1)& \equiv &{{-t}\over{12\pi^2}}
\left(\alpha\theta_1^3 -3\pi \alpha\theta_1^2 +\beta \pi^2\theta_1
+ \gamma \pi^3\right), \\
&=&{{-t\alpha}\over{12\pi^2}}
(\theta_1 - \pi)(\theta_1 -\theta_1^{-})(\theta_1 - \theta_1^{+}),
\end{eqnarray}
where 
%%%%%%%%%
\begin{equation}
\theta_1^{\pm}
=\pi\left(1\pm {{N-1}\over{\sqrt{N^2-3N+3}}}\right).
\end{equation}
Let us note that 
$G(\theta_1)$ is independent of $\bar{m}$ and that
$F(\theta_1)=G(\theta_1)$,
of course, reproduces the equation (\ref{eqevend}).
\par
%%%%%%%%%
We study the behaviors of
the intersections 
of $F(\theta_1)=G(\theta_1)$ with respect to
the scale $\bar{m}$ and the parameter $t$.
We first note that the number of the intersections of the functions
$F(\theta_1)$ and $G(\theta_1)$ is either one or three.
The Higgs VEV is also written, from Eq.(\ref{eqevenu}) after
using the formula (\ref{formulau}), as
%%%%%%%%%%%
\begin{equation}
\vbar^2={1\over{2\pi -\theta_1}}
\left({{-A\alpha}\over{12\pi^2}}\right)
(\theta_1 -\pi)(\theta_1 - \theta_1^-)(\theta_1 - \theta_1^+)\geq 0
\end{equation}
for $\pi \leq \theta_1 \leq 2\pi$.
\par
%%%%%%%%%
It may be useful here to study the matrix $H$ in this case,
which is given by a
$2\times 2$ matrix, as in the previous case (\ref{hessian2}).
We can show that 
the determinant of $H$ is evaluated as
%%%%%%%%
\begin{eqnarray}
{\rm det}~H&=&2\vbar^2(\mbar)\left(-3(1+{{t\alpha}\over
{24\pi^2}})\theta_1^2
+12\pi(1+{{t\alpha}\over {48\pi^2}})\theta_1-{{t\beta}\over {24}}+\mbar^2
-12\pi^2 \right)\nonumber\\
&=&2\vbar^2(\mbar)
\left({{-1}\over 2}\right)
{\del\over{\del\theta_1}}\left(F(\theta_1)-G(\theta_1)\right).
\label{det}
\end{eqnarray} 
We observe that the stability of
the solutions to the equation $F(\theta_1)=G(\theta_1)$
is controlled by the sign of
${\del\over{\del\theta_1}}\left(F(\theta_1)-G(\theta_1)\right)$.
It is also useful to know that
$F(\theta_1=\pi)=G(\theta_1=\pi)$ $=$ $0$ at $\bar{m}=\pi$ and
%%%%%%%%
\begin{equation}
{\del\over{\del\theta_1}}\left(F(\theta_1)-G(\theta_1)\right)
\bigg|_{\mbar=\pi,~~\theta_1=\pi}
=\left\{\begin{array}{lll}
\leq 0 & \mbox{for} & t\geq 48\pi^2 {{N-1}\over N},\\[0.3cm]
> 0   &  \mbox{for} & t < 48 \pi^2 {{N-1}\over N}.
\end{array}\right.
\end{equation}
This observation implies that the intersections of the functions
$F(\theta_1)$ and $G(\theta_1)$ for $t\ge 48\pi^2\frac{N-1}{N}$
and $t<48\pi^2\frac{N-1}{N}$ have different behavior.
We also obtain that
%%%%%%%%%%%
\begin{equation}
F\left(\theta_1=2\pi\pm {\mbar\over{\sqrt 3}}\right)=
\mp {4\over{3\sqrt{3}}}\mbar^3,
\end{equation}
where $\theta_1=2\pi\pm {\mbar\over{\sqrt 3}}$
are the solutions
to $\del F(\theta_1)/\del\theta_1 =0$.
Since
%%%%%%%%%
\begin{equation}
{{\del F(\theta_1)}\over
{\del\bar{m}}
}=4~\mbar(2\pi -\theta_1),
\end{equation}
the function $F(\theta_1)$ increases (decreases) as $\bar{m}$
increases for fixed $\theta_1$ with $\theta_1<2\pi$ ($\theta_1>2\pi$).
Note that $G(\theta_1)$ is independent of $\bar{m}$.
\par
%%%%%%%%%%%%
One can, now, draw the graphs of $F(\theta_1)$ and $G(\theta_1)$ for various
$\mbar$ and $t$ and understand the behavior of the
%intersection of $F(\theta_1)=G(\theta_1)$.
intersections of $F(\theta_1)$ and $G(\theta_1)$.
In Fig.4, we depict
the case $t> 48\pi^2 {{N-1}\over N}$. In the figure, the solution
corresponding to the coexisting phase is denoted by $\theta_c$.
Likewise, in Fig.5, we depict the
case $t<48\pi^2 {{N-1}\over N}$, and the solution
for the coexisting phase is denoted by $\theta_c$.
The other 
solutions give negative determinants of $H$, so that they are
unstable against small fluctuations. We observe from Eq.(\ref{det}) that
the solution $\theta_c$
in Figs.$4$ and $5$ gives a
positive determinant of $H$
and hence the solution, if any, is stable.
It is important to note that
for any $t$, the intersections
in the region of 
$\pi \leq \theta_1 \leq 2\pi$
tend to disappear as the scale $\mbar$ becomes smaller and smaller.
%%%%%%%%%%
\newpage
%%%%%%%%%%%%% BIBLIOGRAPHY %%%%%%%%%%%%%%%%%%%%%%%

%%%%%%%%%%%%%%%%%%%%%%%%%%%%
\newpage
%%%%%%%%%%%%%%%%%%NEW FIGS%%%%%%%%%%%%%%%%%
%%%%%%%%%%%%%%%%%%%%%%%%%%%%%%%%%%%%%%%%%%%
\begin{figure}
\begin{center}
\includegraphics[width=0.8\linewidth]{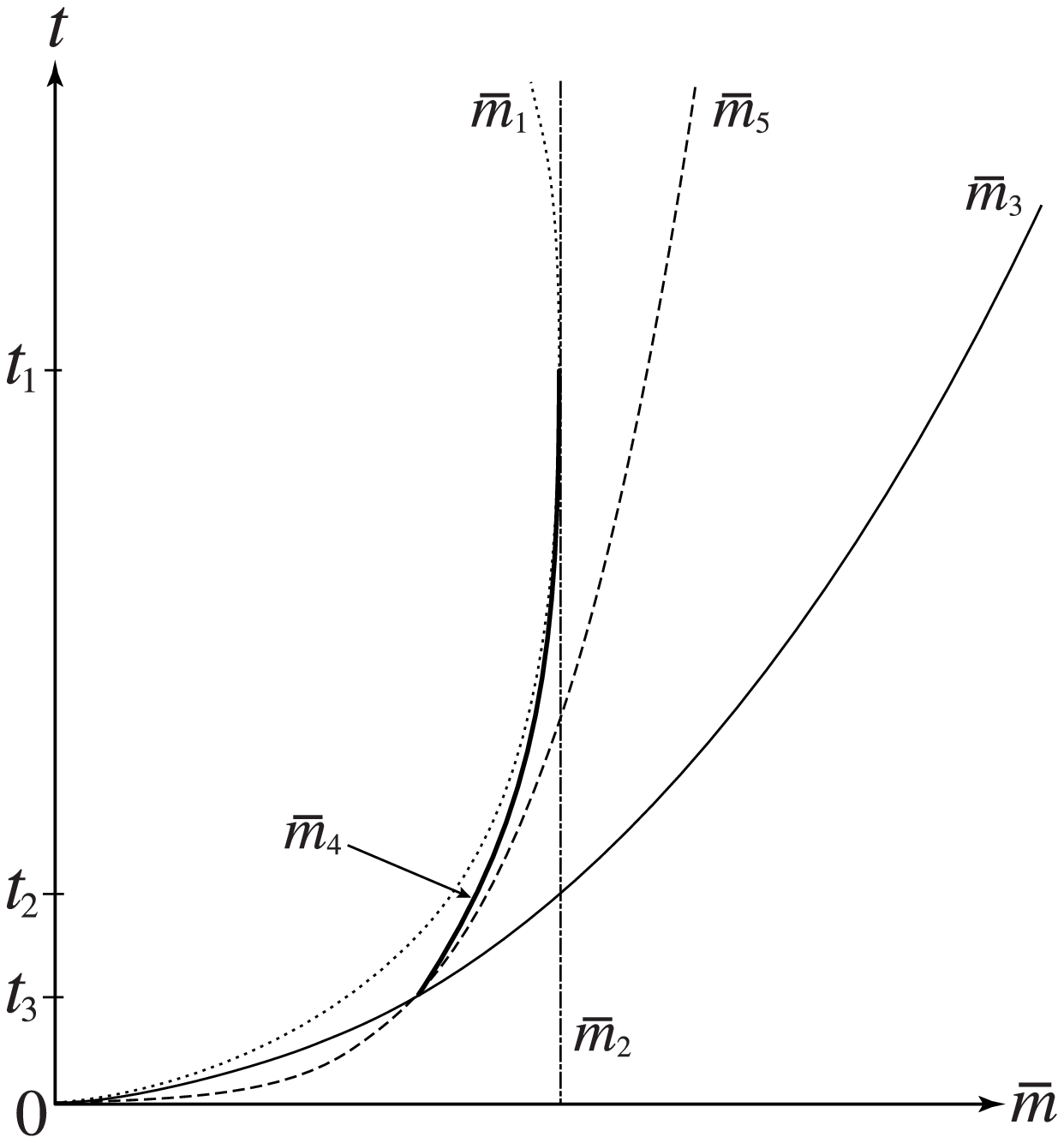}
\end{center}
\caption{
Various critical scales $\bar{m}_i$ $(i=1,\cdots, 5)$ are drawn for the
$SU(N)$ gauge-Higgs model with the fundamental fermion and Higgs fields
for $N=$ odd, 
where $t_1 = \frac{48\pi^2 (N-1)^3}{N(N^2 -3)}$,
$t_2 = \frac{24\pi^2 (N-1)^3}{N^2 (2N-3)}$ and
$t_3 = \frac{24\pi^2 (N-1)^3 (N^2 -N-1)}{N^3(2N-3)^2}$.
}
\end{figure}
%%%%%%%%%%%%%%%%%%
\newpage
%%%%%%
\begin{figure}
\begin{center}
\includegraphics[width=0.8\linewidth]{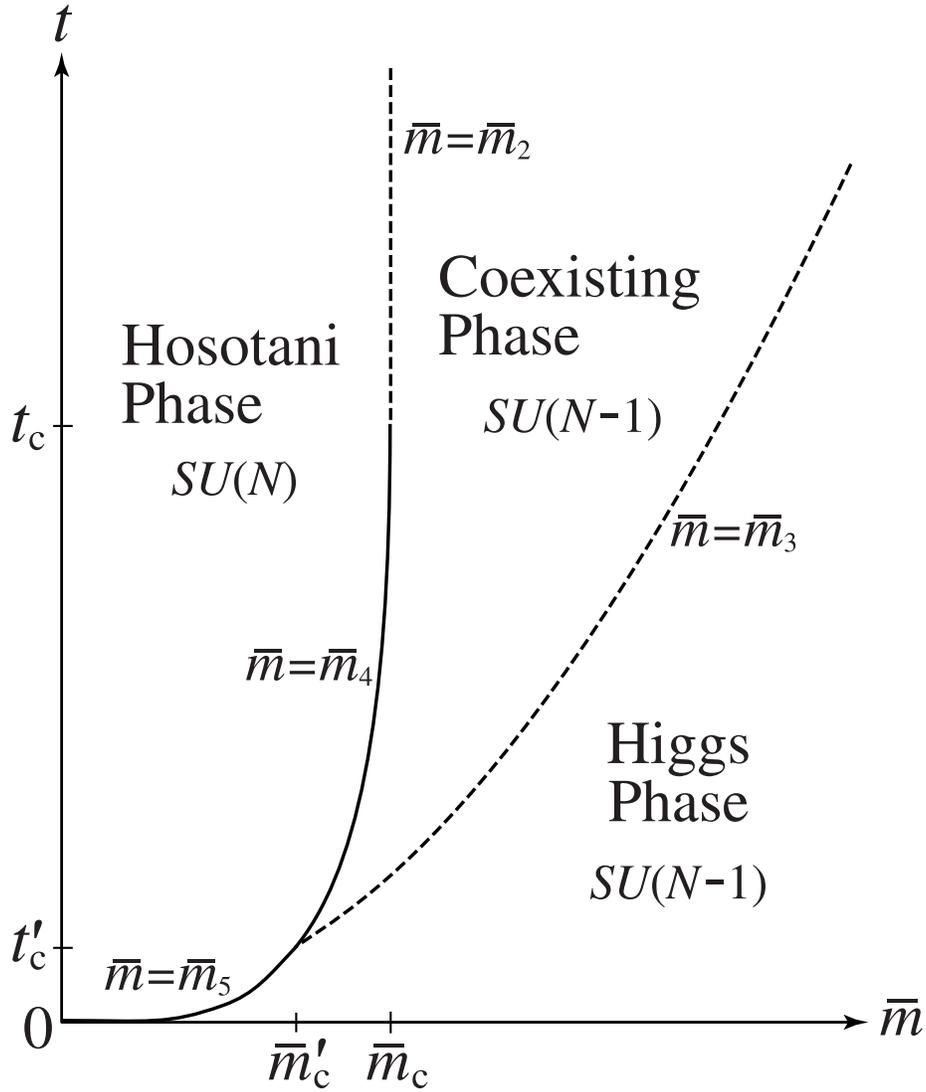}
\end{center}
\caption{
Phase diagram of the $SU(N)$ gauge-Higgs model with the fundamental
fermion and Higgs fields for $N=$ odd,
where $t_c = \frac{48\pi^2 (N-1)^3}{N(N^2 -3)}$,
$t_{c}^{\prime} = \frac{24\pi^2 (N-1)^3 (N^2 -N-1)}{N^3(2N-3)^2}$,
$\mbar_c =\frac{\pi(N-1)}{N}$ and
$\mbar_{c}^{\prime} = \frac{\pi(N-1)}{N}\sqrt{\frac{N^2-N-1}{N(2N-3)}}$.
The solid and dashed curves denote the first- and second-order
phase transitions, respectively.
In each phase, the residual gauge symmetry is shown.
}
\end{figure}
%%%%%%%%%%%%%%%%%%
\newpage
%%%%%%
\begin{figure}
\begin{center}
\includegraphics[width=0.8\linewidth]{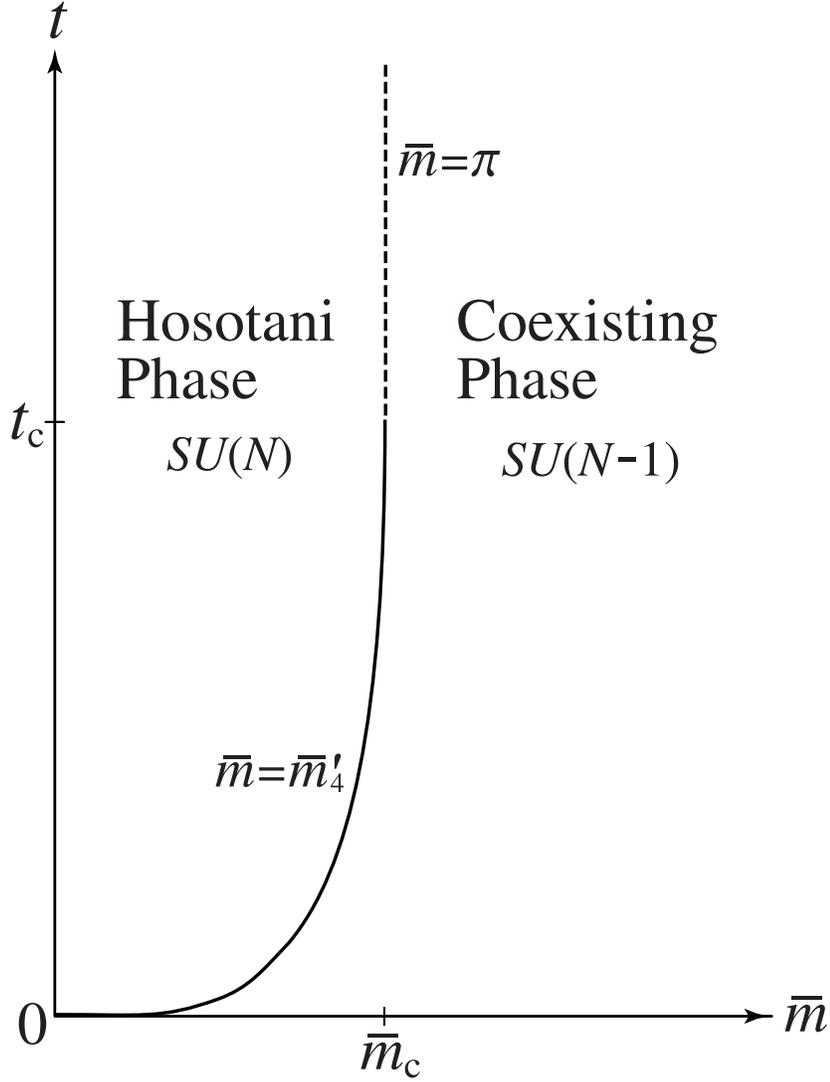}
\end{center}
\caption{
Phase diagram of the $SU(N)$ gauge-Higgs model with the fundamental
fermion and Higgs fields for $N=$ even,
where $t_c = \frac{48\pi^2(N-1)}{N}$ and $\mbar_c =\pi$.
The solid and dashed curves denote the first- and second-order
phase transitions, respectively.
In each phase, the residual gauge symmetry is shown.
}
\end{figure}
%%%%%%%%%%%%%%%%%
\newpage
\begin{figure}
\begin{center}
\includegraphics[width=0.8\linewidth]{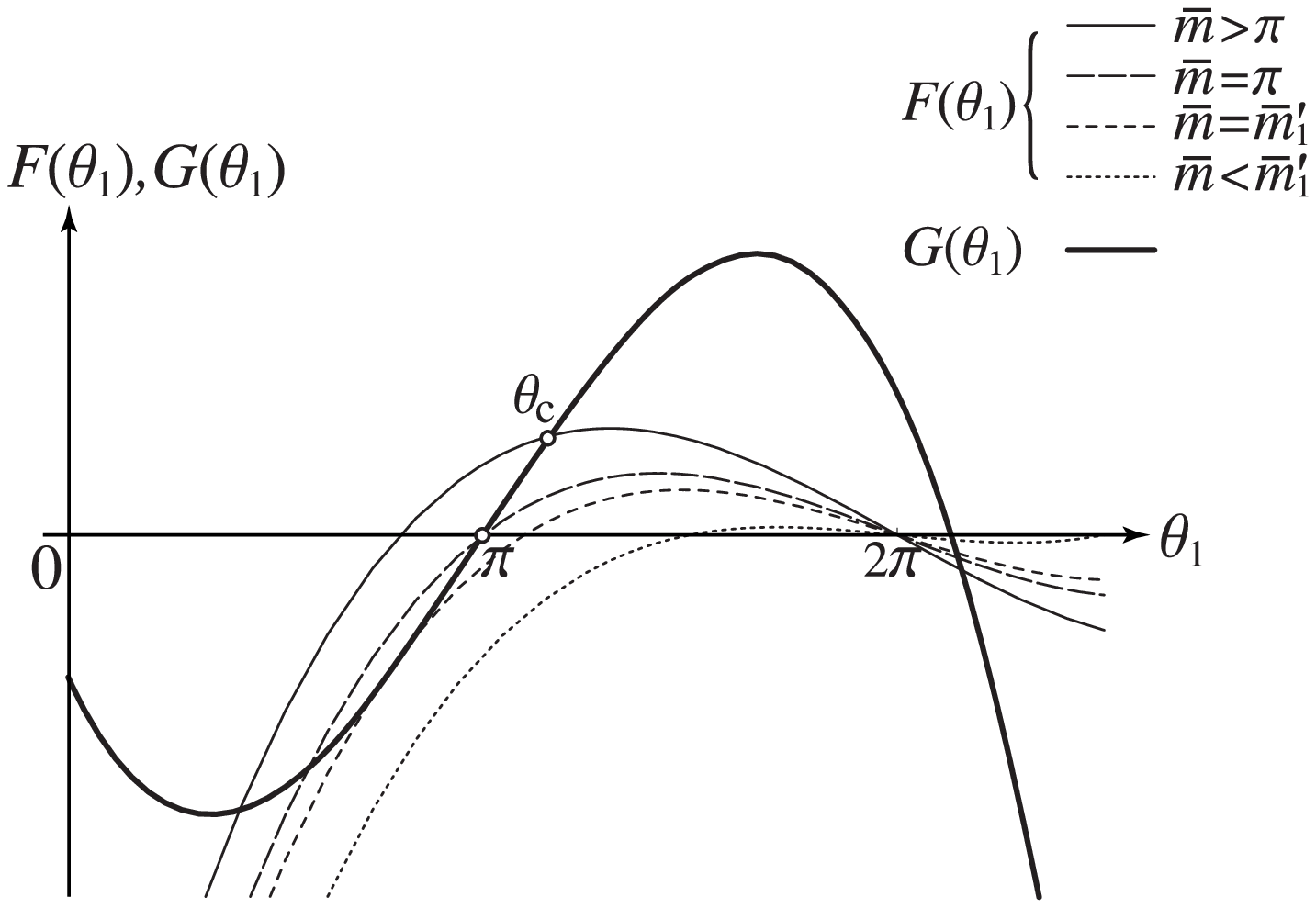}
\end{center}
\caption{
Curves of the function $F(\theta_1)$ and $G(\theta_1)$ for
$t > 48\pi^2 \frac{N-1}{N}$.
The thick curve denotes $G(\theta_1)$, and other thin curves
denote $F(\theta_1)$ for various typical values of $\bar{m}$.
The intersection marked by \lq\lq $\circ$", which lies in the range
of $\pi \le \theta_1 \le 2\pi$ for $\bar{m} \ge \pi$,
corresponds to the type III solution of $\theta_1 = \theta_c$.
}
\end{figure}
%%%%%%%%%%%%%
\begin{figure}
\begin{center}
\includegraphics[width=0.8\linewidth]{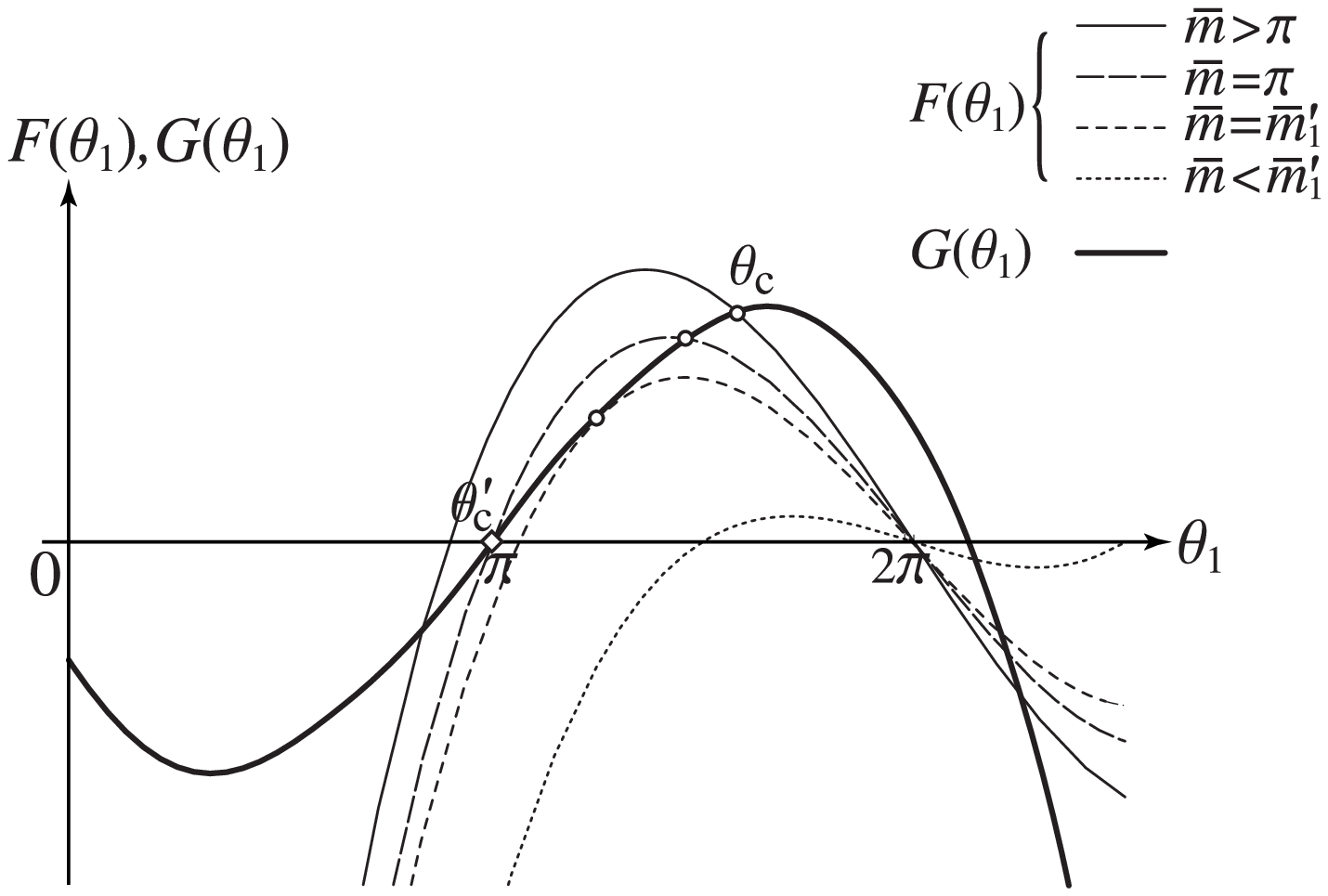}
\end{center}
\caption{
Curves of the function $F(\theta_1)$ and $G(\theta_1)$ for
$t < 48\pi^2 \frac{N-1}{N}$.
The thick curve denotes $G(\theta_1)$, and other thin curves
denote $F(\theta_1)$ for various typical values of $\bar{m}$.
The intersection marked by \lq\lq $\circ$", which lies in the range
of $\pi \le \theta_1 \le 2\pi$ for $\bar{m} \ge \bar{m}_1^{\prime}$,
corresponds to the type III solution of $\theta_1 = \theta_c$.
The other intersection marked by \lq\lq $\diamond$",
which appears in the range of $\pi \le \theta_1 \le 2\pi$
for $\mbar_1'\leq \mbar \le \pi$, corresponds to the
unstable solution of $\theta_1=\theta_c^{\prime}$.
}
\end{figure}
%%%%%%%%%%%%%%%%%%%

\begin{thebibliography}{99}
%%%%%%
\bibitem{SUSY}
M. Sakamoto, M. Tachibana, K. Takenaga,
\PLB{458}{231}{99}, \hepth{9902070};
\PTPM{104}{633}{00}, \hepth{9912229};
in {\it Proceedings of the 30th International Conference
on High Energy Physics, 2000}, edited by C. S. Lim, T. Yamanaka 
(World Scientific, Singapore, 2001),
\hepth{0011058}.
%%%%%%
\bibitem{translation}
M. Sakamoto, M. Tachibana, K. Takenaga, \PLB{457}{33}{99},
\hepth{9902069}.
%%%%%%
\bibitem{O(N) model}
K. Ohnishi, M. Sakamoto, \PLBM{486}{179}{00}, \hepth{0005017}.
\\
H. Hatanaka, S. Matsumoto, K. Ohnishi, M. Sakamoto,
\PRDM{63}{105003}{01}, \hepth{0010283};
in {\it Proceedings of the 30th International Conference
on High Energy Physics, 2000}, edited 
by C. S. Lim, T. Yamanaka (World Scientific, Singapore, 2001),
\hepth{0010041}.
%%%%%%
\bibitem{monopole}
S. Matsumoto, M. Sakamoto, S. Tanimura,
\PLBM{518}{163}{01}, \hepth{0105196}.
\\
M. Sakamoto, S. Tanimura,
\PRDM{65}{065004}{02}, \hepth{0108208}.
%%%%%% Sherk-Schwaltz (Takenaga)
\bibitem{SStakenaga}
K. Takenaga,
\PLB{425}{114}{98}, \hepth{9710058};
\PRD{58}{026004}{98}, {\bf D61} (2000), 129902(E), \hepth{9801075};
\PRDM{64}{066001}{01}, \hepth{0105053};
\PRDM{66}{085009}{02}, \hepth{0205173}.
%%%%%% Hosotani mech. on S^1
\bibitem{Hosotani}
Y. Hosotani, 
\PLB{126}{309}{83}, \ANN{190}{233}{89}.
%%%%%% Hosotani mech. on S^
\bibitem{HosotaniS2}
Y. Hosotani, \PLB{129}{193}{83}.
%%%%%% Hosotani mech. on Torus
\bibitem{HosotaniT2}
J.E. Hetrik, C.L. Ho,
\PRD{40}{4085}{89}.
\\
A. McLachlan,
\NPB{338}{188}{89}.
%%%%%% Hosotani mech. and Congruency Class
\bibitem{HP}
A. Higuchi, L. Parker,
\PRD{37}{2853}{88}.
%%%%%
%\\
%A.T. Davies, A. McLachlan,
%\NPB{317}{237}{89}.
%%%%%%% Hosotani mech. and Finite Temperature
\bibitem{Hoso-ft}
K. Shiraishi,
\PTP{77}{975}{87}.
%%%%
%C.L. Ho, Y. Hosotani,
%\NPB{345}{445}{90}.
%%%%%%% Hatanaka-Inami-Lim
\bibitem{HIL}
H. Hatanaka, T. Inami, C.S. Lim,
\MPL{13}{2601}{98}, \hepth{9805067}.
%%%%%%%matter contents
\bibitem{Pattern-Matter}
H. Hatanaka, \PTP{102}{407}{99}, \hepth{9905100}.
%%%%%%% 
\bibitem{Pattern-BC}
C.L. Ho, Y. Hosotani, \NPB{345}{445}{90}.
%%%%%%% Orbifold
\bibitem{orbifold}
L. J. Dixon, J. A. Harvey, C. Vafa, E. Witten,
\NPB{261}{678}{85}.
%%%%%
%Nucl.\ Phys.\ B {\bf 261}, 678 (1985).
%%%%%%%
%\bibitem{orbifold-GUT}
%%%%%%% Hosotani mech on Orbifold
\bibitem{Hoso-orbi}
M. Kubo, C.S. Lim, H. Yamashita,
\MPLM{17}{2249}{02}, \hepph{0111327}.
\\
N. Haba, M. Harada, Y. Hosotani, Y. Kawamura,
\NPBM{657}{03}{169}, \hepph{0212035}.
%%%%%%% Previous
\bibitem{previous} 
H. Hatanaka, K. Ohnishi, M. Sakamoto, K. Takenaga,
\PTPM{107}{1191}{02},
\hepth{0111183}.
%%%%%%% finitetemperature
\bibitem{finitetemperature}
L. Dolan, R. Jackiw, \PRD{9}{3320}{74}.\\
S. Weinberg, \PRD{9}{3357}{74}.
%%%%%%% finiteradius
\bibitem{finiteradius}
L.H. Ford, T. Yoshimura, \PLA{70}{89}{79}.\\
D.J. Toms, \PRD{21}{928}{80}; \PRD{21}{2805}{80}.\\
G. Denardo, E. Spallucci, \NPB{169}{514}{80};
\NCA{A58}{243}{80}.
%%%%%%% adjointmodel
\bibitem{adjointmodel}
A.T. Davies, A. McLachlan, \NPB{317}{237}{89}. 
%%%%%%% 
\end{thebibliography}
\end{document}